\documentclass[fleqn,usenatbib]{mnras}

\usepackage{newtxtext,newtxmath}

\usepackage[T1]{fontenc}
\usepackage{ae,aecompl}

\usepackage{soul}
\renewcommand\hl[1]{#1} 

\usepackage{graphicx}	
\usepackage{amsmath}	

\title[Simulated stripped nuclei and UCDs]{Contribution of stripped nuclei to the ultracompact dwarf galaxy population in the Virgo Cluster}

\author[R. J. Mayes et al.]{Rebecca J. Mayes,$^{1}$\thanks{E-mail: r.mayes@uq.net.au}
Michael. J. Drinkwater,$^{1}$
Joel Pfeffer,$^{2}$ Holger Baumgardt,$^{1}$  
\newauthor Chengze Liu,$^{3}$ Laura Ferrarese,$^{4}$ Patrick C{\^o}t{\'e},$^{4}$ and Eric W. Peng$^{5,6}$
\\
$^{1}$School of Mathematics and Physics, University of Queensland, Brisbane, QLD 4072, Australia\\
$^{2}$Astrophysics Research Institute, Liverpool John Moores University, 146 Brownlow Hill, Liverpool L3 5RF, UK\\
$^{3}$Department of Astronomy, School of Physics and Astronomy, and Shanghai Key Laboratory for Particle Physics and Cosmology, \\
Shanghai Jiao Tong University, Shanghai 200240, P. R. China\\
$^{4}$Herzberg Astronomy and Astrophysics Research Centre, National Research Council of Canada, Victoria, BC V9E 2E7, Canada\\
$^{5}$Department of Astronomy, Peking University, Beijing China 100871\\
$^{6}$ Kavli Institute for Astronomy and Astrophysics, Peking University, Beijing China 100871\\
}

\date{Accepted XXX. Received YYY; in original form ZZZ}

\pubyear{2019}

\begin{document}
\label{firstpage}
\pagerange{\pageref{firstpage}--\pageref{lastpage}}
\maketitle

\begin{abstract}
We use the hydrodynamical EAGLE simulation to predict the numbers, masses and radial distributions of tidally stripped galaxy nuclei in massive galaxy clusters,  and compare these results to observations of ultra-compact dwarf galaxies (UCDs) in the Virgo cluster. We trace the merger trees of galaxies in massive galaxy clusters back in time and determine the numbers and masses of stripped nuclei from galaxies disrupted in mergers. The spatial distribution of stripped nuclei in the simulations is consistent with those of UCDs surrounding massive galaxies in the Virgo cluster. Additionally, the numbers of stripped nuclei are consistent with the numbers of M > 10\textsuperscript{7}~\(\textup{M}_\odot\) UCDs around individual galaxies and in the Virgo cluster as a whole. The mass distributions in this mass range are also consistent. \hl{We find that the numbers of stripped nuclei surrounding individual galaxies correlates better with the stellar or halo mass of individual galaxies than the total cluster mass.} We conclude that most high mass (M >  10\textsuperscript{7}~\(\textup{M}_\odot\)) UCDs are likely stripped nuclei. It is difficult to draw reliable conclusions about low mass (M < 10\textsuperscript{7}~\(\textup{M}_\odot\)) UCDs because of observational selection effects. We additionally predict that a few hundred stripped nuclei below a mass of 2~$\times$~10\textsuperscript{6}~\(\textup{M}_\odot\) should exist in massive galaxies that will overlap in mass with the globular cluster population. Approximately 1-3 stripped nuclei in the process of forming also exist per massive galaxy.

\end{abstract}




\section{Introduction}

Ultra-compact dwarf galaxies (UCDs) were first discovered in spectroscopic surveys of the Fornax cluster \citep{Hilker1999, Drinkwater2000} and have since been discovered in other clusters \citep{Mieske2004, Mieske2007a, has2005, Madrid2010, Misgeld2011, caso2013, Zhang2015}, in galaxy groups \citep{Evstig2007a,DaRocha2011, Madrid2013} and around isolated galaxies 
\citep{Hau200}. 
UCDs appear as objects intermediate between globular clusters and dwarf galaxies having absolute magnitudes  -14.0~mag < M\textsubscript{v} < -10~mag \citep{Voggel2016} and half-light radii of 7~pc < r\textsubscript{h} < 100~pc \citep{Mieske2008}. They have central velocity dispersions similar to dwarf galaxies, of approximately 20 < $\sigma$\textsubscript{0} < 50~km s\textsuperscript{-1}, giving dynamical masses of approximately 2~$\times$~10\textsuperscript{6}
 to 10\textsuperscript{8}~\(\textup{M}_\odot\)
 \citep{has2005, Hilker2008, Mieske2008, Mieske2013}. They typically have old stellar populations, with ages of at least 8 Gyr \citep{Chilingarian2011, Janz_2015}. The method by which UCDs form is a matter of some debate, and there are a variety of competing formation theories. One scenario contends that they are the high-mass end of the globular cluster mass-function around galaxies with rich GC systems \citep{Mieske2002, Mieske2012}, possibly formed from the merger of many globular clusters in star cluster complexes \citep{Kroupa1998, Fellhauer2002, brun2011, brun2012}. The alternate scenario is that they are the compact nuclei of dwarf galaxies that underwent galaxy threshing leaving only the nucleus remaining \citep{Bassino1994, Bekki2001, Bekki2003, Drinkwater2003, Goerdt2008, Pfeffer2013, Pfeffer2014}. There is a growing body of evidence that neither formation scenario
is responsible for the full UCD population and that both the main scenarios make up some portion of the UCD population \citep{Mieske2006, brodie2011, Chilingarian2011, DaRocha2011,Norris2011,Pfeffer2014, Pfeffer2016}. Aside from a few distinct objects that contain central supermassive black holes or feature extended star formation histories \citep[e.g.][]{Seth2014, Norris2015, Ahn2017, Ahn2018} it is difficult to distinguish between the different formation scenarios since they predict similar internal UCD properties. Therefore, determining what role each mechanism plays in the formation of UCDs requires predictions from simulations on how each process may contribute to the formation of UCD populations.

Stripped nuclei are a likely explanation for at least some percentage of the UCD population for a variety of reasons. In structure and properties, UCDs bear many resemblances to the compact nuclei of galaxies: They overlap the luminosity distribution of nuclei \citep{Drinkwater2004}, and follow a similar size-luminosity distribution while being approximately 2.2 times larger than galaxy nuclei at the same luminosity \citep{Evstig2008A}. They have similar internal velocity dispersions \citep{Drinkwater2003}, follow dwarf elliptical nuclei on the colour-magnitude diagram \citep{Cote2006, Evstig2008A, brodie2011}, and have similar metallicities to dwarf galaxies, while lying above the metallicity-luminosity relation \citep{Chilingarian2011, Francis2012, Spengler2017, Zhang2018}. This would be expected in the case of tidal stripping as it would cause luminosity to decrease while metallicity remains constant.  \hl{\mbox{\citet{Janz_2015}} found a transition in UCD stellar populations at \mbox{M~=~2~$\times$~10\textsuperscript{7}~\(\textup{M}_\odot\)}, which they interpreted as the point at which UCDs transition from being composed of a mixture of stripped nuclei and globular clusters to being composed primarily of stripped nuclei.} Some UCDs have been found to have stellar halos which might be the remains of the original galaxies \citep{Drinkwater2003, has2005, Chilingarian2008, Evstig2008A, Chiboucas2011}. Other UCDs show evidence of tidal tail-like features \citep{Voggel2016}, and a UCD in the Virgo cluster, M59-UCD3, has a tidal stream pointing towards it which is potentially the remnant of its tidally stripped galaxy \citep{LiuPeng2015}. Irregular objects have been found which may be dwarf galaxy nuclei in the process of being tidally stripped \citep{Richtler2005, brodie2011, Jennings2015}. Tidal stripping is a confirmed origin for several UCDs \citep{Seth2014, Norris2015} due to the indirect evidence of a supermassive black hole. The nearby nucleated dwarf galaxy Sagittarius which is undergoing tidal stripping around the Milky Way is the closest example of a possible UCD in formation \citep{Ibata1994, Carretta2010}. \hl{Tidal stripping is also a possible formation mechanism for compact elliptical (cE) galaxies \mbox{\citep[e.g.][]{Huxor_2011}}}.

Pioneering simulations established that galaxy stripping can produce objects with similar properties to UCDs \citep{Bassino1994, Bekki2001, Bekki2003, Pfeffer2013}. Several simulations have been carried out for UCD formation in static potentials \citep{Bekki2003, Goerdt2008, Thomas2008}, and suggest that the radial distributions of UCD populations should be centrally concentrated. The models predict too few UCDs at large radii, however, and static models have limitations, such as being unable to account for UCD formation within smaller sub-clusters that subsequently merged with the larger one, as well as an inability to account for time-varying potentials. Triaxial potentials resulting in box or other chaotic orbits are also possible for dwarf galaxies \citep{Pfeffer2013} and may provide more accurate models of UCD formation. \citet{Pfeffer2014} and \citet{Pfeffer2016} carried out the first investigations of the contribution of galaxy stripping to the UCD population using cosmological simulations of galaxy formation. By modelling the formation of stripped nuclei in the Millennium II simulations \citep{Boylan2009}, combined with a semi-analytic model for galaxy formation \citep{Guo2011}, it was found that at most 10 per cent of UCDs in the Fornax cluster could have formed by tidal stripping. However, semi-analytic models of galaxy formation have some limitations. The lack of a baryonic component for galaxies affects the time taken for them to undergo stripping, and the numbers of galaxies stripped. Modelling UCD formation with hydrodynamical simulations may provide more accurate results.

In this paper, we use the EAGLE simulation of galaxy formation and evolution \citep{Crain_2015, Schaye2015} to simulate our new method of stripped nuclei formation and predict the numbers, masses and distributions of stripped nuclei. We then compare these properties to observations. We aim to determine whether the radial distributions, numbers and masses of UCDs in large galaxy clusters can be explained by tidal stripping of galactic nuclei, whether the UCD population is primarily dependant on central galaxy mass or host cluster mass, and also whether transitional objects, galaxies in the process of being stripped, exist in the simulations. Throughout the paper, the objects formed are referred to as stripped nuclei as they resemble both GCs and UCDs and more than one formation channel may contribute to UCD formation.

This paper has the following organization. The method by which we identified stripped nuclei in the EAGLE simulations is described in Section 2. The results of our research are presented in Section 3. We discuss the implications of our work for models of UCD formation in Section 4, and a summary of our results is given in Section 5.

\section{Methods and observations} \label{method}

In this section, we give an overview of the EAGLE simulations used to identify stripped nuclei by tracking the most tightly bound star particles of galaxies in the simulation. We also give an overview of the observations of UCDs and the methods used for comparison.  

\label{sec:maths} 

\subsection{Overview of simulations}

Hydrodynamical simulations of galaxy formation can, in principle, predict the distribution of galaxies much more accurately than semi-analytic simulations due to the addition of baryonic particles. By directly accounting for the presence of baryonic matter, fluid motions can be calculated as well as gravity. This will impact the timescale on which the gas in satellite galaxies undergoes ram-pressure stripping and the stellar component is stripped through tidal forces, and therefore influence the number of galaxies stripped in the simulation. Additionally, the presence of baryonic matter allows for properties of galaxies such as stellar and black hole mass to be determined directly from the baryonic particles in the simulation. In this work, we use the state-of-the-art hydrodynamical EAGLE simulations \citep{Schaye2015, Crain_2015}. The EAGLE simulations are well tested and reproduce many properties of evolving galaxy populations. Feedback from supernovae and black holes is calibrated to reproduce the z = 0 galaxy mass function, galaxy sizes and black hole masses \citep{Schaye2015}. The simulations also reproduce the evolution of the galaxy mass function \citep{Furlong2015} and galaxy sizes \citep{Furlong2017}, galaxy luminosities and colours \citep{Trayford2015}, cold gas properties \citep{Lagos2015, Crain_2015}, and largely reproduce the cosmic star formation rate density and specific SFR-galaxy mass relation \citep{Furlong2015}.

The largest of the EAGLE simulations has a box side length of 100 comoving Mpc, large enough to contain ten thousand galaxies of the mass of the Milky Way or larger, and is made up of a total of 6.8 billion particles. Baryonic particles in the simulation have an initial mass of 1.81~$\times$~10\textsuperscript{6}~\(\textup{M}_\odot\)
and dark matter particles a mass of 9.70~$\times$~10\textsuperscript{6}~\(\textup{M}_\odot\),
 meaning galaxies with as low a stellar-mass as 1~$\times$~10\textsuperscript{8}~\(\textup{M}_\odot\) are resolved by more than 50 particles. 
 We use three forms of data from the EAGLE simulation in this paper:
\begin{enumerate}
  \item [\textbf{Online Database}]
  \item Halo and galaxy data for the simulation is stored in an online database \citep{McAlpine2016}. The database can be used to determine information such as galaxy masses and merger trees for use in the analysis of tidally stripped nuclei without relying on the more cumbersome particle data.
  \item [\textbf{Raw Particle Data}]
  \item The individual particle data can be downloaded from the EAGLE website. The raw particle data are required to determine the locations of the stripped nuclei, but we did the bulk of our analysis using the online database, to reduce the amount of information that must be processed.
  \item [\textbf{Linking Database}]
  \item We linked the online database and raw particle data through a database we created that connects the ID of each galaxy in the online database to the ID of its most bound star particle in the particle data.
\end{enumerate}

\subsection{Simulated cluster selection}
As our objective is to compare the properties of tidally stripped nuclei in the simulations to those of observed UCDs in the Virgo cluster, we begin by selecting clusters in the most massive of the EAGLE simulations, RefL0100N1504, which has a box size of 100 comoving Mpc containing 2~$\times$~1504\textsuperscript{3} particles. The most massive cluster in this simulation has a Friend-of-Friends (FoF) mass of 6.4~$\times$~10\textsuperscript{14}~\(\textup{M}_\odot\) and $\mathcal{M}$\textsubscript{200} = 1.87~$\times$~10\textsuperscript{14}~\(\textup{M}_\odot\). We further select 6 other clusters in the simulation with $\mathcal{M}$\textsubscript{200} above 1~$\times$~10\textsuperscript{14}~\(\textup{M}_\odot\) for subsequent analysis. Properties of the seven clusters are outlined in Table \ref{tab:clusters_analysed}.

The Friend-of-Friends (FoF) mass is determined by linking spatially connected structure. A linking length is defined, and every nearby particle that has a distance shorter than that length is connected \citep{Davis1985}. EAGLE has a defined linking length of 0.2 times the mean interparticle separation. The critical mass or $\mathcal{M}$\textsubscript{200} is the total mass within the $\mathcal{R}$\textsubscript{200} radius, which is the physical radius within which the cluster density is 200 times the critical density of the Universe. The critical mass is generally considered to be a more observationally comparable property than the FoF mass, but it does not take into account non-spherical distributions since it uses a single radius. This can be a problem for the most massive structures since they are often not fully virialized and/or may be composed of several massive substructures. Similar issues can occur when considering observations of clusters such as Virgo. 

We will use both these masses in this paper, and when considering $\mathcal{M}$\textsubscript{200}, we will only include the stripped nuclei that are found within $\mathcal{R}$\textsubscript{200}. For comparisons with Virgo we will work primarily with the most massive simulated cluster which has a 
similar luminosity function to Virgo. In cases where we directly compare the masses of our simulated clusters to Virgo we will use the Virgo masses M = 5.5~$\times$~10\textsuperscript{14}~\(\textup{M}_\odot\) \citep{Durrell2014}, which was obtained by combining the masses of subclusters within Virgo, making it comparable to the FoF mass, and a critical mass of $\mathcal{M}$\textsubscript{200} = 4.2~$\times$~10\textsuperscript{14}~\(\textup{M}_\odot\) \citep{McLaughlin1999}. 
\begin{table*}
	\centering
	\caption{List of clusters analysed from the EAGLE simulation}
	\label{tab:clusters_analysed}
	\begin{tabular}{lccc} 
		\hline
		Cluster ID & Friend-of-Friends Mass (\(\textup{M}_\odot\)) & Critical Mass (\(\textup{M}_\odot\)) & Most Massive Galaxy (\(\textup{M}_\odot\))\\
		\hline
		28000000000000 & 6.42~$\times$~10\textsuperscript{14} & 1.87		$\times$ 10\textsuperscript{14} & 4.65 $\times$ 10\textsuperscript{11}\\
		28000000000001 & 6.22 $\times$ 10\textsuperscript{14} & 3.73 $\times$ 10\textsuperscript{14} & 3.53 $\times$ 10\textsuperscript{11}\\
		28000000000002 & 3.76 $\times$ 10\textsuperscript{14} & 3.00 $\times$ 10\textsuperscript{14} & 2.99 $\times$ 10\textsuperscript{11}\\
		28000000000003 & 3.48 $\times$ 10\textsuperscript{14} & 3.07 $\times$ 10\textsuperscript{14} & 4.48 $\times$ 10\textsuperscript{11}\\
		28000000000004 & 2.50 $\times$ 10\textsuperscript{14} & 1.96 $\times$ 10\textsuperscript{14} & 2.05 $\times$ 10\textsuperscript{11}\\
		28000000000005 & 2.31 $\times$ 10\textsuperscript{14} & 1.98 $\times$ 10\textsuperscript{14} & 3.73 $\times$ 10\textsuperscript{11}\\
		28000000000006 & 2.05 $\times$ 10\textsuperscript{14} & 1.30 $\times$ 10\textsuperscript{14} & 2.37 $\times$ 10\textsuperscript{11}\\
		\hline
	\end{tabular}
\end{table*}

\subsection{Identifying stripped nuclei}
UCDs are too compact to be fully resolved in the EAGLE simulations, but the EAGLE particle masses of 1.81~$\times$~10\textsuperscript{6}~\(\textup{M}_\odot\) for baryonic particles is similar to the mass range of observed UCDs (2~$\times$~10\textsuperscript{6}~\(\textup{M}_\odot\) to 10\textsuperscript{8}~\(\textup{M}_\odot\)). Therefore, we define the central, most bound star particle (MBP) of a galaxy before its merger as the nucleus of the galaxy. \hl{We use that MBP to track the position of the nucleus following the merger, through to redshift zero}. \hl{Star particles are not affected by hydrodynamic forces, and so will behave in a similar way gravitationally to nuclear star clusters}, meaning they can be used to model nuclei behaviour. We did not use the other types of particles found in the simulation to track nuclei because they may behave differently than nuclear star clusters: i.e., a gas particle disappearing from the simulation due to accretion or a black hole particle experiencing mergers. \hl{The exact star particle that is most bound can vary from snapshot to snapshot, so we tested this method by repeating some of our analysis with the 5th most bound particle. This was found not to affect our results.}

To create a sample of candidate galaxies that could be disrupted to become stripped nuclei, we carried out the following steps. Note that we designed the process to create a sample of possible progenitor galaxies in the simulation and a sample of their most bound particles. The numbers of galaxies or stripped nuclei remaining after each step are listed in Tables \ref{tab:28000000000000} and \ref{tab:21242350}.

\begin{enumerate}
  \item First, for a cluster in the simulation, we define the massive galaxies in the cluster at z = 0 that will have potentially disrupted smaller galaxies in the past. Galaxies in the simulation are defined as potential disrupting galaxies if they have a stellar-mass greater than 10\textsuperscript{7}~\(\textup{M}_\odot\), as this mass approaches the lower limit at which galaxies can be defined in the EAGLE simulation.
  \item In order to identify stripped nuclei, we traced back the merger trees of these massive galaxies to find mergers involving any progenitor galaxies with stellar mass >  10\textsuperscript{7}~\(\textup{M}_\odot\).  \citet{Sanchez2019} found that at a stellar mass of 10\textsuperscript{7}~\(\textup{M}_\odot\) approximately 30 per cent of galaxies are nucleated and have a mean nuclear star cluster to galaxy mass ratio of 1.7 per cent. Therefore, our stripped nuclei sample will be unreliable below a mass of $\approx$ 1.7~$\times$~10\textsuperscript{5}~\(\textup{M}_\odot\). We will also lose several objects above this mass due to scatter in the nuclear star cluster to galaxy mass ratio relation. As we primarily focus on M > 2~$\times$~10\textsuperscript{6}~\(\textup{M}_\odot\) objects, this should not affect our results.
  \item Galaxies in the merger tree may appear in multiple snapshots before merger, so we reduce the merger tree to unique galaxies. This was done by selecting the progenitor galaxy immediately before its merger, and determining the particle ID of the most bound star particle in this snapshot.  
  \item We then used the particle data to determine properties for the most bound particles, defined as galaxy nuclei, such as their position at z = 0.
\end{enumerate}

The next step was to determine whether the nuclei of the progenitor galaxies could have survived to z = 0 and become potential UCDs. We consider that a stripped nucleus is formed in a merger if the following conditions are satisfied:

\begin{enumerate}
  \item The merger between the progenitor and the central galaxy was a minor merger rather than a major one. For a merger to be minor, the stellar mass ratio between the two galaxies must be smaller than 1/4 \citep{Qu2017}. During major mergers instead of one galaxy being stripped by a more massive one, both galaxies will be highly disrupted \citep{Casteels2014}. 
  \item The time the stripped nucleus has been orbiting its central galaxy is shorter than its dynamical friction timescale. We calculated the dynamical friction timescale with equation 7-26 from \citet{Binney1987}, modified with an eccentricity function as defined in Appendix B of \citet{Lacey1993}. To calculate the eccentricity, we first determine the circular velocity of the particle around the central galaxy from the \hl{virial} velocity and radius of the halo. The angular momentum of the particle around the central galaxy is then calculated from its velocity and position relative to the central galaxy. Next, the energy of the particle is calculated from the circular velocity, particle velocity and radius around the central galaxy. This energy is then used to calculate the radius the particle would have if it were on a circular orbit. The angular momentum for the circular orbit is then calculated from the circular radius and velocity. Next, the eccentricity is determined by taking the ratio of the true angular momentum and the circular orbit angular momentum. Finally, the dynamical friction is calculated using the eccentricity function \hl{\mbox{\citep[eq. B4 from][]{Lacey1993}}}, circular radius, circular velocity and mass of the stripped nucleus (see Section~\ref{section:mass}).
  
  Stripped nuclei with dynamical friction timescales shorter than the orbital time will have spiralled inwards and merged with the central galaxy before the final (z = 0) snapshot.
\end{enumerate}

\subsection{Nucleation fraction and nucleus mass }
\label{section:mass}

UCDs are most commonly defined as having masses greater than 2~$\times$~10\textsuperscript{6}~\(\textup{M}_\odot\) \citep[e.g.][]{Mieske_2008}, so our analysis will primarily focus on objects above this mass. Two crucial aspects of our study that will impact the numbers and masses of our simulated UCDs are the fraction of stripped galaxies that are nucleated and the masses of the nuclei within those galaxies.  
\subsubsection{Nucleation fraction}

We base our estimate of the number of galaxies that are nucleated on Figure 2 from \citet{Sanchez2019}, which plots the fraction of nucleated galaxies in the Virgo, Fornax and Coma clusters. In the Virgo Cluster, the galaxy nucleation fraction peaks at approximately 90 per cent for galaxies with stellar mass $\approx$ 10\textsuperscript{9}~\(\textup{M}_\odot\) and linearly declines for more and less massive galaxies, on a logarithmic scale. The nucleation fraction reaches zero at approximately M\textsubscript{*} = 10\textsuperscript{11}~\(\textup{M}_\odot\) for high mass galaxies and 5~$\times$~10\textsuperscript{5}~\(\textup{M}_\odot\) for low mass galaxies. When considering the nucleation fraction for galaxies, we primarily work with fractions of stripped nuclei rather than choosing a random sample of stripped nuclei in each mass range.

\subsubsection{Nucleus mass}

The masses of the resulting stripped nuclei can be estimated from the original progenitor galaxy's stellar mass. We base our nuclei mass estimates on Figure 9. from \citet{Sanchez2019}, which plots the ratio of nuclear star cluster to galaxy mass as a function of galaxy stellar mass. The mass ratio has a minimum of approximately 0.36 per cent for galaxies with stellar mass 3~$\times$~10\textsuperscript{9}~\(\textup{M}_\odot\) and then increases for more and less massive galaxies. 
To determine the masses of the stripped nuclei, we assign each one a mass randomly chosen from a log-normal mass function for the nucleus-to-galaxy mass ratio with a mean chosen by applying linear fits to the upper and lower mass ranges of Figure 9 in \citet{Sanchez2019} and a log-normal standard deviation of 0.4 dex. 

Tidal stripping for the nuclear cluster is not taken into account. This may cause us to slightly overestimate the masses of the resulting stripped nuclei if they lose mass during the merger, although, the nucleus is known to lose little mass during the stripping process \citep{Bekki2001, Bekki2003}. There is also the possible case of the nucleus retaining some mass from the host galaxy halo so the stripped nucleus would be more massive than the nuclear cluster itself \citep{Pfeffer2013, LiuPeng2015,Voggel2016}, which would cause us to underestimate the masses of the resulting stripped nuclei

\subsection{Simulation corrections} \label{simcorr}

There are several methodology issues related to the use of the EAGLE simulations. These are induced by the limitations of the simulations or our methods of tracking stripped nuclei.

One issue that relates to studying UCD formation in EAGLE is the simulation mass resolution, induced by using a single particle to represent a galactic nucleus. It is possible for the particle to be ejected during the merger before the merger is complete and therefore not accurately follow the path of the stripped nucleus. Our choice of the most bound star particle before merging as the nucleus makes this less probable.

Another issue is that galaxies in the merger tree occasionally exhibit unstable behaviour during mergers \citep{Qu2017}, which complicates the determination of the major-minor ratio, and the mass of the resulting stripped nucleus. Occasionally two merging galaxies will appear to change masses radically between snapshots. This is due to the \textsc{subfind} \citep{Springel2001, Dolag2009} routine identifying particles attached to the central galaxy as being attached to the progenitor galaxy, resulting in an apparent ``see-saw" change in central and progenitor galaxy masses between snapshots. Fewer than 5 per cent of mergers exhibit this behaviour, but it can lead to an overestimate of stripped nuclei masses and numbers. To combat this, when determining mass ratios, we take the mass ratio from the snapshot with the maximum central galaxy stellar mass in the five snapshots before merger.

Due to stripping during a merger event, a progenitor galaxy in the snapshot immediately before a merger may have lost stellar mass found in earlier snapshots. Therefore for progenitor galaxies that do not exhibit switching behaviour, we take the maximum stellar mass in all snapshots before merger. In galaxies which do switch masses, we take the stellar mass from the snapshot where the central galaxy stellar mass is at a maximum, which may lead to a slight underestimate of stellar mass for the progenitor galaxy in these cases.

The \textsc{subfind} algorithm occasionally erroneously identifies small dense stellar regions within galaxies as separate distinct objects \citep{Schaye2015}. These objects are not genuine galaxies, and so we discarded progenitor galaxies flagged as spurious in the EAGLE database.

Around 4 per cent of merging galaxies appear to be massive galaxies without any progenitors and an extremely low (< 10\textsuperscript{7}~\(\textup{M}_\odot\)) dark matter content. These galaxies are anomalous self-bound regions of the host galaxy from a past merger the host galaxy experienced \citep{Wetzel2009}. We excluded these galaxies as potential progenitors by removing galaxies with dark matter content < 10\textsuperscript{7}~\(\textup{M}_\odot\).

Because of the presence of diffuse stars and the possibility of particles being miss-assigned to the wrong halo it is recommended that the stellar mass of a galaxy be measured using a spherical aperture in EAGLE \citep{McAlpine2016}. We use an aperture \hl{radius} of 30~kpc, which is recommended due to being suited for comparisons to observations \citep{Schaye2015}. 

\subsection{Observational Data} 
\label{obsdata}
To determine the contribution that stripped nuclei make to the UCD population, we made comparisons with observations of UCDs in the Virgo cluster. 
We made single galaxy radial distribution comparisons with UCDs taken from the Next Generation Virgo Cluster Survey \citep{Ferrarese2012, Liu2015}. The survey finds 92 confirmed UCDs around M87, the galaxy closest to the centre of the potential for the Virgo cluster, and 28 and 23 around the other massive galaxies M49 and M60 respectively. The UCDs were defined as having projected half-light radii 11 < r\textsubscript{h} < 100~pc. The magnitude range of UCDs for this study 
was 18.5 $\leqslant$ g $\leqslant$ 21.5~mag (-12.7 $\leqslant$ M\textsubscript{g} $\leqslant$ -9.7, Virgo distance of 16.5~Mpc, distance modulus = 31.087 \citep{Mei2007}). 

With a mass-to-light ratio of 2.15 \(\textup{(M/L)}_\odot\) applied \citep{Voggel_2019} 
this magnitude range converts roughly to a mass range of 1.6~$\times$~10\textsuperscript{6}~\(\textup{M}_\odot\) $\lesssim$ M $\lesssim$ 2.58~$\times$~10\textsuperscript{7}~\(\textup{M}_\odot\). UCDs close to the center of galaxies may have been missed due to light saturation. 

The mass-size relation of UCDs reaches 10~pc at a magnitude of  M\textsubscript{v}~$\approx$~-11~mag, or mass M~$\approx$~10\textsuperscript{7}~\(\textup{M}_\odot\) \citep{Norris2011}. Because of the lower radius limit of 11 pc used to take this sample the UCDs at masses below $\approx$~10\textsuperscript{7}~\(\textup{M}_\odot\) are likely to be undersampled. Therefore when comparing this sample to our simulation data, we will primarily work with stripped nuclei above this mass. This limit will also result in a smaller number of UCDs with M > 10\textsuperscript{7}~\(\textup{M}_\odot\) being undersampled. 

\begin{figure}
	\includegraphics[width=\columnwidth]{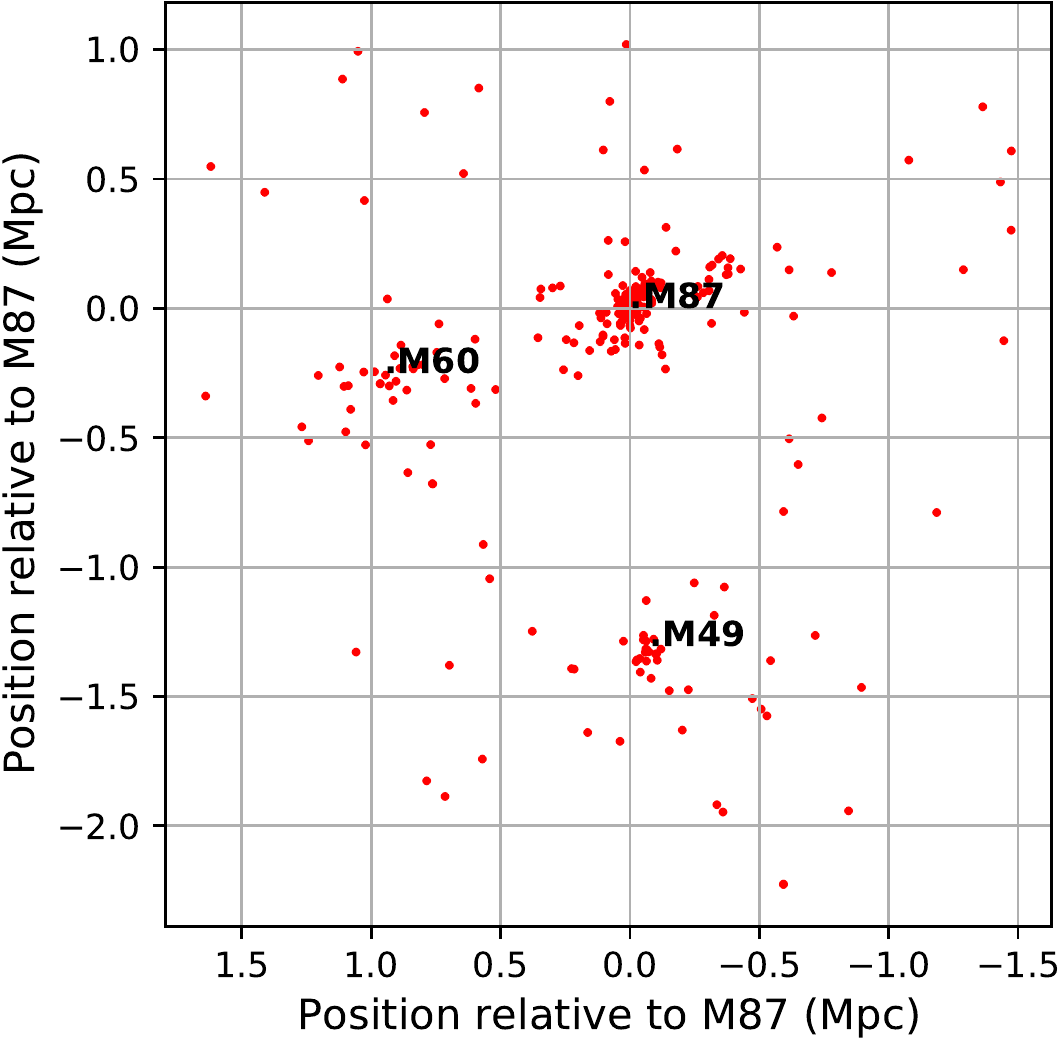}
    \caption{Projected, observed, distribution of 243 UCD candidates in the Virgo cluster, described in Section \ref{obsdata}. \citet{liu2020generation}.}
    \label{fig:virgoscatter}
\end{figure}

Our analysis with the full simulated clusters was made using a new sample of 243 UCD candidates in the Virgo cluster \citet{liu2020generation}. The completeness level of the sample is over 90 per cent and is selected using the same range in r\textsubscript{h} as the \citet{Liu2015} UCDs, with magnitude g~$\leqslant$~21.5~mag (mass range M~$\geqslant$~1.6~$\times$~10\textsuperscript{6}~\(\textup{M}_\odot\)). The sample is subject to the same radius-mass under-sampling as the single galaxy comparison. Fig.~\ref{fig:virgoscatter} shows the distribution of these UCDs on the sky, with the position converted from right ascension and declination to megaparsecs.

\begin{table*}
	\centering
	\caption{Most massive EAGLE Cluster: ID = 28000000000000}
	\label{tab:28000000000000}
	\begin{tabular}{lr} 
		\hline
		Step & Number\\
		\hline
		Galaxies in cluster with stellar mass >  1 $\times$  10\textsuperscript{7} \(\textup{M}_\odot\) & 678 galaxies\\
		Post stripped nuclei processing: Galaxies with stripped nuclei & 313 galaxies\\
		Post stripped nuclei processing: Galaxies with stripped nuclei M >  2~$\times$~ 10\textsuperscript{6}~\(\textup{M}_\odot\) & 111 galaxies\\
		\hline
	\end{tabular}
\end{table*}

\begin{table*}
	\centering
	\caption{Most massive galaxy in Cluster ID = 28000000000000: Galaxy ID = 21242350}
	\label{tab:21242350}
	\begin{tabular}{lr} 
		\hline
		Step & Number\\
		\hline
        Number of unique galaxies in merger tree & 22056 Galaxies  \\
        Galaxies with stellar mass > 1 $\times$  10\textsuperscript{7} \(\textup{M}_\odot\)& 1238 Galaxies  \\
        Number of galaxies that merged via minor mergers  & 1122 Galaxies\\
        Number of stripped nuclei that did not spiral in due to dynamical friction & 1080 Stripped Nuclei\\ 
        Number of stripped nuclei M >  2 $\times$  10\textsuperscript{6} \(\textup{M}_\odot\) & 145 Stripped Nuclei\\
        Non-anomalous galaxies & 118  Stripped Nuclei\\
        Fraction nucleated & 80 Stripped Nuclei\\
		\hline
	\end{tabular}
\end{table*}

When considering the number of stripped nuclei found around simulated galaxies, we can consider both the ``merger-tree" and the ``aperture" \hl{samples}. The merger tree \hl{sample} is taken directly from disrupted galaxies found in the merger tree of the central galaxy and is a direct representation of the number of satellite galaxies stripped by the central galaxy. This \hl{sample} is the most reliable for determining the number of stripped nuclei directly formed by progenitors disrupted by a galaxy, however, it may not be reliable when making comparisons with observations. This method will not account for stripped nuclei clustering around a massive galaxy that may have been stripped by other smaller galaxies close to it, and it will also include merger tree nuclei that may no longer be associated with the central galaxy, that would not be seen in observations.

A more accurate method for comparing simulated stripped nuclei to observations of UCDs is to take an aperture sample of stripped nuclei found close to the galaxy. This is done by sampling all the stripped nuclei within a defined radius around the galaxy. Thus the aperture sample will include stripped nuclei from other galaxies close in projected distance to the central galaxy, and exclude merger tree stripped nuclei that may be located some distance away. This provides a more accurate comparison to observations, as observed samples of UCDs are based on projected distance to massive galaxies. However, it has the limitation of some stripped nuclei possibly being ``double-counted" if we are analysing multiple galaxies close to each other and the apertures we are using for the multiple galaxies overlap. \hl{ Throughout the paper the radius used for the aperture sample differs depending on the variable measured but is chosen to be consistent with observations. The distance the stripped nuclei sample extends to is typically larger than the 30 kpc aperture radius EAGLE galaxy stellar masses are measured within.}

For both the merger tree and the aperture \hl{samples}, we use a projected distribution, for the sake of more accurate comparisons to observations.

\section{Results}

In this section, we present results from the analysis of the EAGLE simulations described in Section \ref{method}. Where possible, we work with fractions of stripped nuclei rather than randomly choosing a number of stripped nuclei to satisfy the progenitor galaxy nucleation fractions, with direct number quotations using fractions, and plots primarily working with a random selection.

\subsection{Properties of most massive simulated cluster}
\begin{figure}
	\includegraphics[width=\columnwidth]{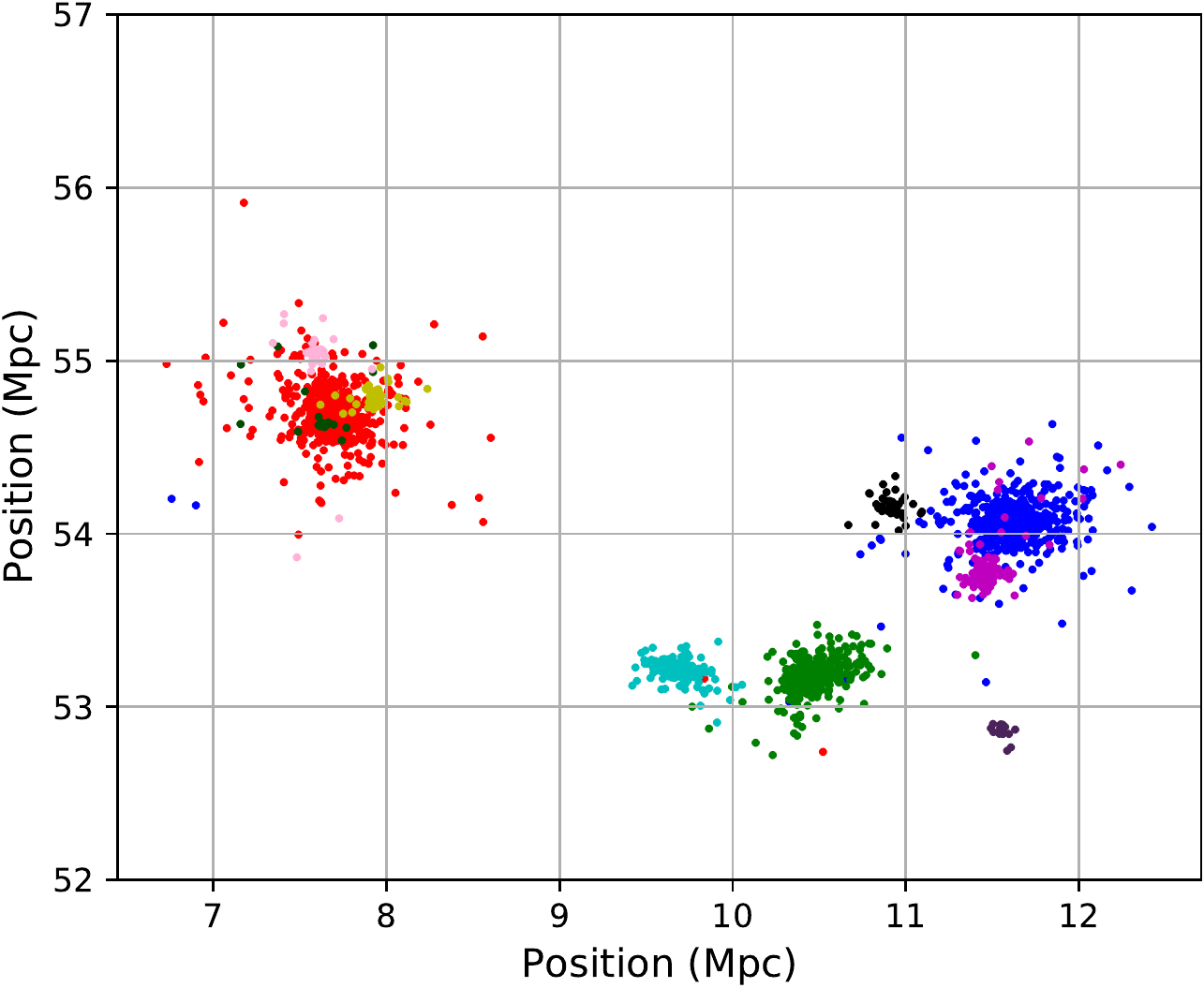}
    \caption{Projected distribution of all merger tree stripped nuclei formed by the 10 most massive galaxies in the most massive EAGLE cluster. The stripped nuclei largely cluster around the galaxy their progenitors merged with.}
    \label{fig:10galsproj}
\end{figure}

Fig.~\ref{fig:10galsproj} shows the distribution of all the stripped nuclei from the merger trees of the ten most massive galaxies of the most massive cluster analysed, \hl{projected in the x-y plane of the simulation volume (i.e. a random projection)}. The stellar masses of these galaxies and the number of stripped nuclei they formed via merging progenitors is listed in Table.~\ref{tab:cluster000}. Similarly to Fig.~\ref{fig:virgoscatter}, the stripped nuclei are not spread throughout the cluster but largely grouped around massive galaxies. Several of the stripped nuclei populations for the galaxies appear to overlap in the projected view. We find that this cluster has a total of $400 \pm 53$ stripped nuclei above a mass of 2~$\times$~10\textsuperscript{6}~\(\textup{M}_\odot\). Of these, $51.8 \pm 7.7$ were massive stripped nuclei M~>~1~$\times$~10\textsuperscript{7}~\(\textup{M}_\odot\) and $2.3 \pm 1.1$ were more massive than 1~$\times$~10\textsuperscript{8}~\(\textup{M}_\odot\).

\begin{table*}
	\centering
	\caption{The 10 most massive galaxies of galaxy cluster ID = 28000000000000, and their merger tree stripped nuclei}
	\label{tab:cluster000}
	\begin{tabular}{llcc} 
		\hline
		Galaxy ID & Stellar mass & No. M >  2 $\times$  10\textsuperscript{6} \(\textup{M}_\odot\) & No. M > 10\textsuperscript{7} \(\textup{M}_\odot\)\\
		\hline
        21242350 & 4.65 $\times$ 10\textsuperscript{11} & $80 \pm 10$ & $10.8 \pm 3.0$ \\
        21109760 & 4.20 $\times$ 10\textsuperscript{11} & $74 \pm 11$ & $11.3 \pm 2.3$ \\
        18481114 & 3.06 $\times$ 10\textsuperscript{11} & $41.4 \pm 6.0$ & $6.1 \pm 1.8$ \\
        13892596 & 1.86 $\times$ 10\textsuperscript{11} & $17.1 \pm 3.0$ & $3.0 \pm 1.1$ \\
        13921560 & 1.85 $\times$ 11\textsuperscript{11} & $13.9 \pm 2.9$ & $2.2 \pm 1.1$ \\
        13938864 & 1.03 $\times$ 10\textsuperscript{11} & $5.5 \pm 1.5$ & $0.75 \pm 0.48$ \\
        13914718 & 9.61 $\times$ 10\textsuperscript{10} & $7.0 \pm 1.6$ & $1.20 \pm 0.89$ \\
        8092293 & 8.98 $\times$ 10\textsuperscript{10} & $5.1 \pm 1.4$ & $1.00 \pm 0.79$ \\
        8071905 & 8.68 $\times$ 10\textsuperscript{10} & $4.40 \pm 0.34$ & $0.41 \pm 0.56$ \\
        13935854 & 7.81 $\times$ 10\textsuperscript{10} & $1.60 \pm 0.79$ & $0.04 \pm 0.15$ \\
		\hline
	\end{tabular}
\end{table*}

A total of 109 galaxies in the cluster were found to be producers of M~>~2~$\times$~10\textsuperscript{6}~\(\textup{M}_\odot\) stripped nuclei. The massive, central galaxies of the cluster were found to be the primary producers of stripped nuclei. The most massive galaxy (shown in red in Fig.~\ref{fig:10galsproj}) produced $80 \pm 10$ or $19.9 \pm 4.5$ per cent of the stripped nuclei found in the whole cluster, including $20.8 \pm 1.8$ per cent of stripped nuclei above a mass of 1~$\times$~10\textsuperscript{7}~\(\textup{M}_\odot\) and $13.6 \pm 2.5$ per cent of the stripped nuclei above a mass of 1~$\times$~10\textsuperscript{8}~\(\textup{M}_\odot\). The two most massive galaxies in the cluster produced $38.5 \pm 9.0$ per cent of the total M~>~2 $\times$~10\textsuperscript{6}~\(\textup{M}_\odot\) stripped nuclei population of the cluster and $43 \pm 14$ per cent of stripped nuclei above a mass of 1~$\times$~10\textsuperscript{7}~\(\textup{M}_\odot\).

\subsection{Comparisons between the most massive simulated galaxy and the central Virgo cluster galaxy}

\label{212comps}



\begin{figure}
	\includegraphics[width=\columnwidth]{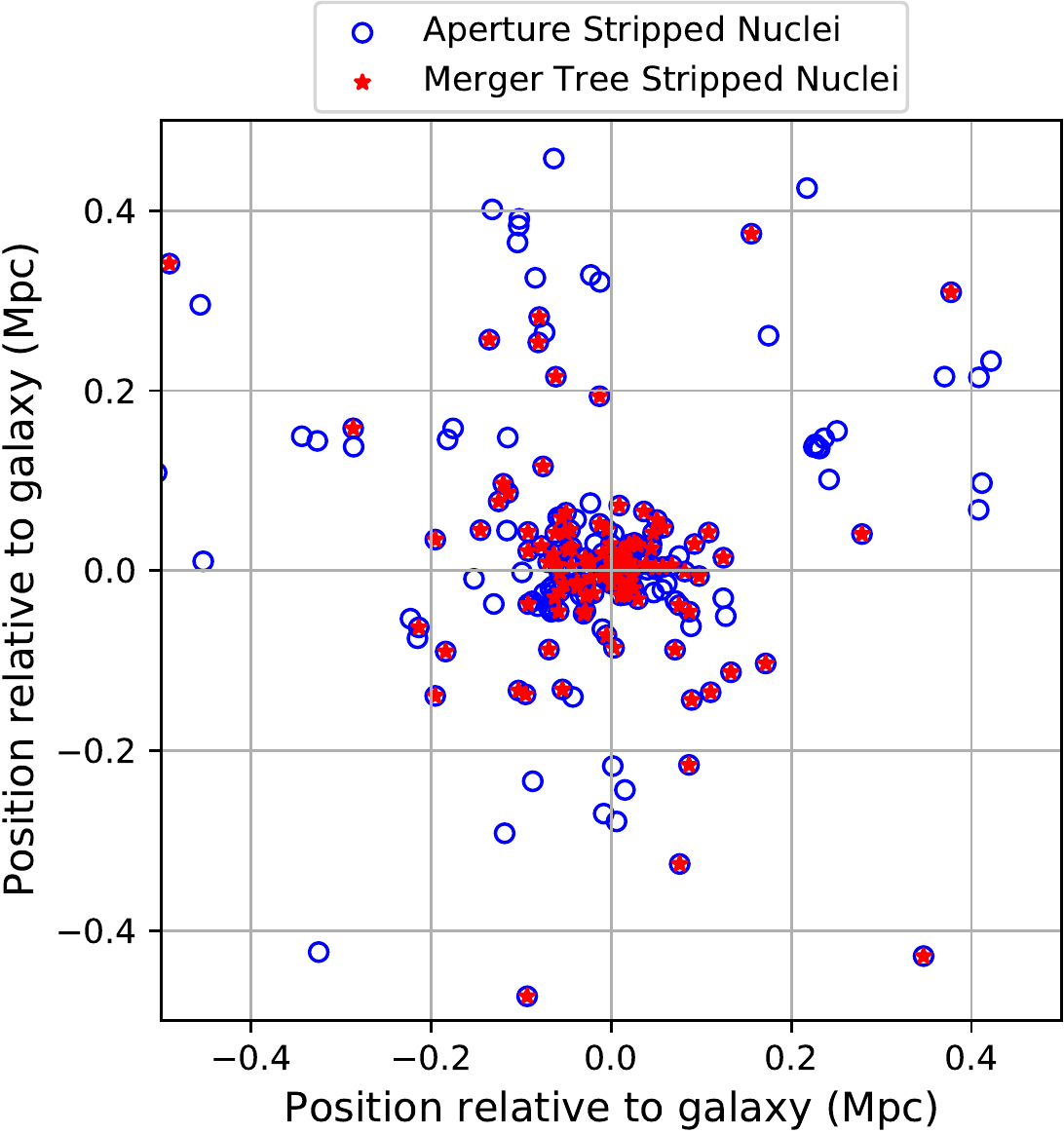}
    \caption{\hl{X-Y plane} projected merger-tree and aperture \hl{sampled} M~>~2~$\times$ 10\textsuperscript{6}~\(\textup{M}_\odot\) stripped nuclei for the most massive galaxy of the simulated cluster, Galaxy ID = 21242350.}
    \label{fig:212scatterplot}
\end{figure}

Fig.~\ref{fig:212scatterplot} depicts the distribution of merger tree and aperture stripped nuclei with mass >~2~$\times$~10\textsuperscript{6}~\(\textup{M}_\odot\) surrounding the most massive galaxy in the most massive cluster in the EAGLE simulation. The merger tree stripped nuclei are represented by red stars and are the nuclei of the progenitor galaxies that were stripped by the massive galaxy. The aperture stripped nuclei are shown with open blue circles and are a mixture of this galaxies merger tree nuclei and stripped nuclei from the merger trees of satellite galaxies.
In both cases, the stripped nuclei cluster strongly around the massive galaxy with over 50 per cent of the nuclei located within 100 kiloparsecs of the central galaxy. Applying the Kolmogorov-Smirnov test to the radial distributions of the merger tree and aperture stripped nuclei returns p~=~0.234, indicating that the two distributions are consistent. 

Table \ref{tab:sn21242350} shows the numbers of stripped nuclei found for both the merger tree and aperture \hl{samples} of this galaxy. The number of M~>~2~$\times$~10\textsuperscript{6}~\(\textup{M}_\odot\) stripped nuclei increases by 50 per cent from the merger tree to the aperture \hl{sample}, while the number of M~>~1~$\times$~10\textsuperscript{7}~\(\textup{M}_\odot\) stripped nuclei increases by 27 per cent. 

This suggests that the numbers of stripped nuclei surrounding galaxies are more sensitive to aperture \hl{sampling} than the radial distributions are. When comparing with observations, we will only use the aperture selection because the observed samples can only use apertures. When comparing different simulated galaxies, we will also consider the merger tree sample. 

\begin{table*}
	\centering
	\caption{Number of stripped nuclei around simulated galaxy ID = 21242350}
	\label{tab:sn21242350}
	\begin{tabular}{llcc} 
		\hline
		Sample & No. M >  2 $\times$  10\textsuperscript{6} & No. M > 10\textsuperscript{7} \(\textup{M}_\odot\) & No. M > 10\textsuperscript{8} \(\textup{M}_\odot\)\\
		\hline
        Merger tree & $80 \pm 10$ & $10.8 \pm 3.0$ & $0.31 \pm 0.37$ \\
        Aperture & $120 \pm 14$ & $13.7 \pm 3.8$ & $0.95 \pm 0.37$ \\
		\hline
	\end{tabular}
\end{table*}
\subsubsection{Radial distributions of stripped nuclei compared with UCDs in the Virgo cluster}

\begin{figure}
	\includegraphics[width=\columnwidth]{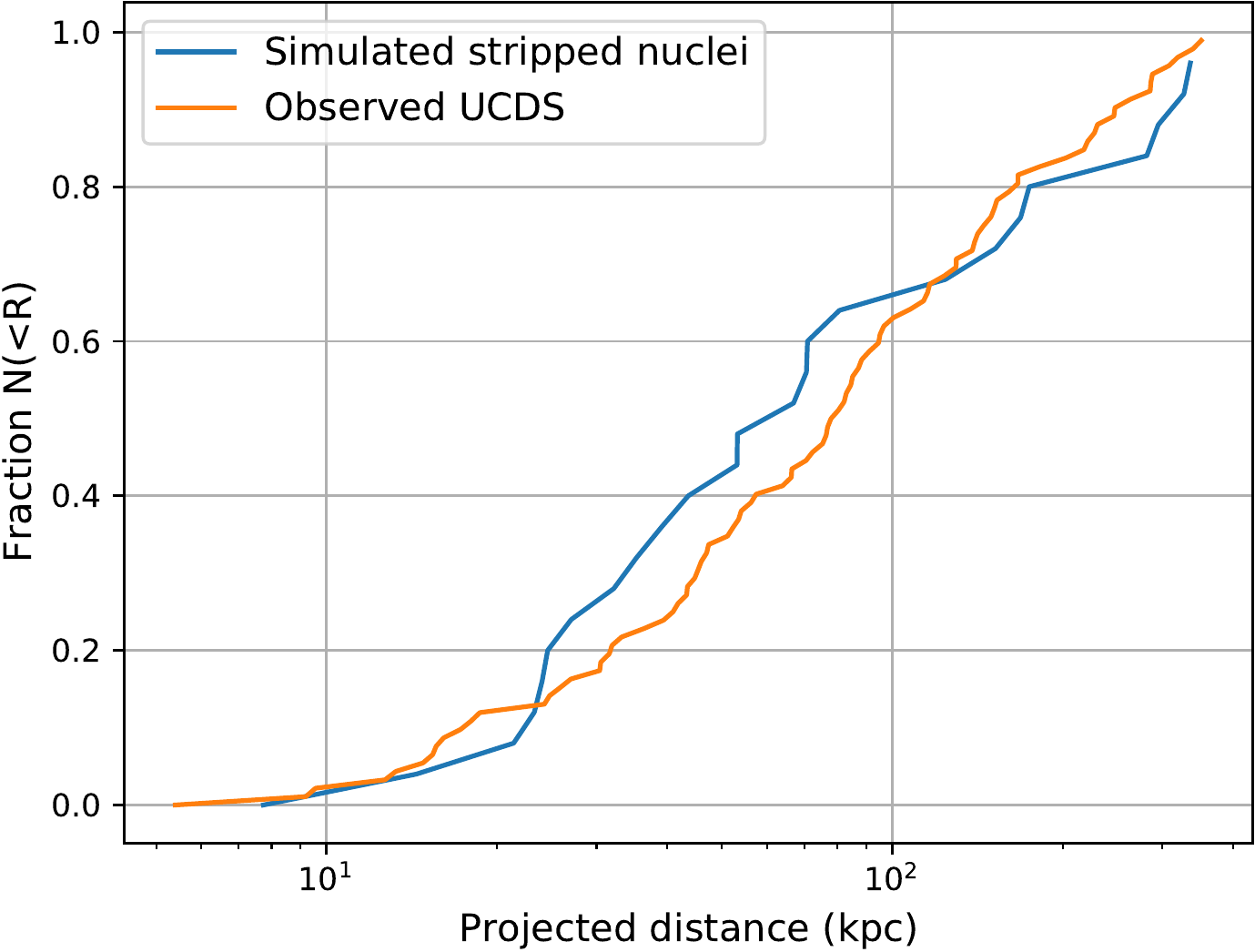}
    \caption{Aperture sampled radial distribution of M~>~1~$\times$~10\textsuperscript{7}~\(\textup{M}_\odot\) stripped nuclei associated with the most massive simulated galaxy of the cluster and UCDs around M87 \citep{Liu2015}. An outer radius limit of 400~kpc was imposed on the simulated distribution to account for observational limitations. The two distributions were found to be consistent (KS test, p~=~0.460.)}
    \label{fig:distnew}
\end{figure}

The next step is to compare the numbers and distributions of our simulated stripped nuclei to observations of UCDs in the Virgo cluster. 
In Fig.~\ref{fig:distnew}, we compare the cumulative radial distributions of the aperture \hl{sampled} distribution of stripped nuclei associated with the most massive galaxy of the simulated cluster with UCDs associated with the dominant galaxy of the Virgo cluster, M87 \citep{Liu2015}. Both galaxies are located at centre of potentials in their respective clusters and have similar masses. Within a 30 kpc radius the simulated galaxy has a stellar mass of 4.65~$\times$~10\textsuperscript{11}~\(\textup{M}_\odot\), while M87 has a stellar mass of 5.5~$\times$~10\textsuperscript{11}~\(\textup{M}_\odot\) \citep{Gebhardt_2009}. An upper radius limit of 400 kpc was imposed on the simulated stripped nuclei to match the observational selection. Visually the two distributions appear to have a strong correlation and applying the Kolmogorov-Smirnov test to these two distributions returns p~=~0.460, indicating that they are consistent.


\subsubsection{Mass distributions of stripped nuclei and UCDs}


We next compare masses of the stripped nuclei from the central most massive galaxy of the most massive cluster to observed UCDs around M87. For both galaxies, we use aperture \hl{samples}, with a 400~kpc aperture, and the Virgo UCDs chosen from \citet{liu2020generation}. Above 1~$\times$~10\textsuperscript{7}~\(\textup{M}_\odot\) the simulated galaxy contains $14 \pm 4$ stripped nuclei, while M87 contains 18. Therefore stripped nuclei are fully consistent with making up the 1~$\times$~10\textsuperscript{7}~\(\textup{M}_\odot\) UCDs that surround M87, when considering a singular central massive galaxy in two similar clusters. 

\begin{figure}
	\includegraphics[width=\columnwidth]{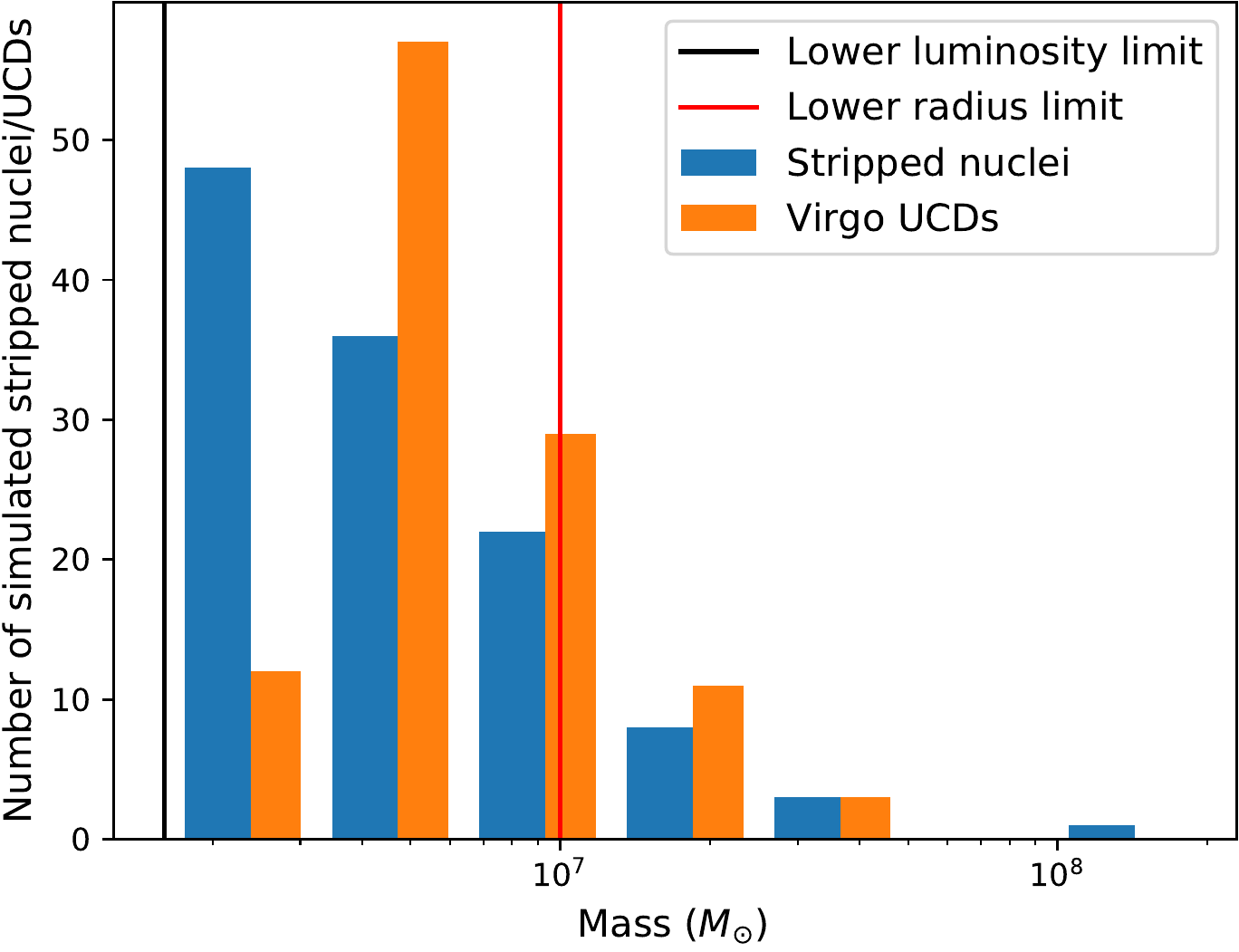}
    \caption{Comparative mass distributions for stripped nuclei from the most massive simulated galaxy and M87 UCDs. The black line indicates the lower luminosity limit for observed UCDs, and the red line the lower radius limit below which UCDs are undersampled, converted to mass limits as described in Section~\ref{obsdata}. The two distributions were consistent above a mass of 1~$\times$~10\textsuperscript{7}~\(\textup{M}_\odot\) (KS test, p~=~0.97). The objects are divided into seven bins from a minimum mass of 1.6~$\times$~10\textsuperscript{6}~\(\textup{M}_\odot\) to a maximum of 2~$\times$~10\textsuperscript{8}~\(\textup{M}_\odot\).}
    \label{fig:212masses}
\end{figure}

Applying the Kolmogorov-Smirnov test to these two distributions for UCDs and stripped nuclei above a mass of 1~$\times$~10\textsuperscript{7}~\(\textup{M}_\odot\) returns p~=~0.97 indicating, that the mass distributions are consistent. 
Fig.~\ref{fig:212masses} plots the comparative mass distributions of both galaxies.


\subsection{Comparisons with Full Virgo Cluster}
We next compare the numbers and distributions of our stripped nuclei sample from the full simulated cluster with observations of UCDs throughout the entire extent of the Virgo cluster.




\subsubsection{Mass distribution comparison}
\begin{figure}
	\includegraphics[width=\columnwidth]{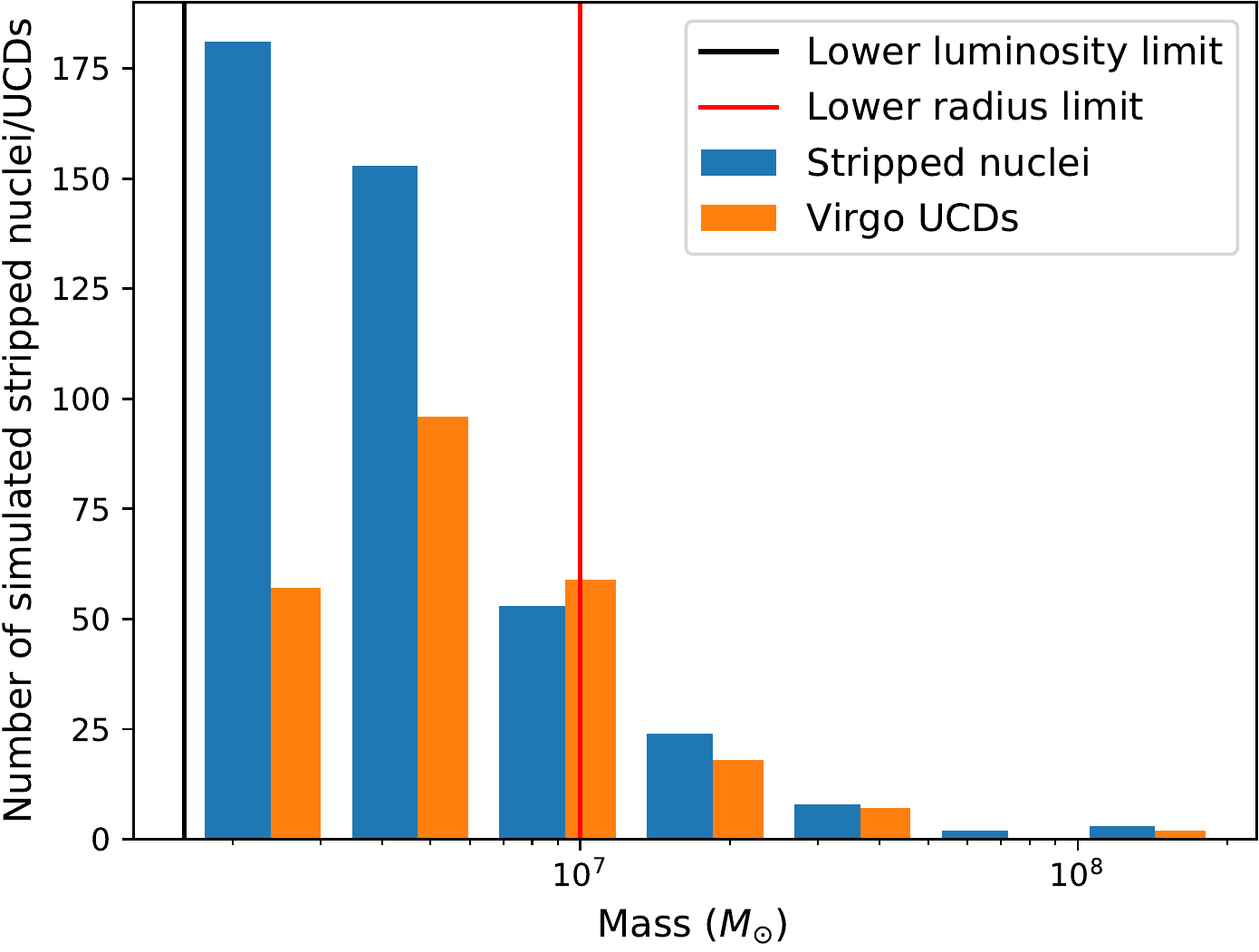}
    \caption{Comparative mass distributions for stripped nuclei in the simulated cluster and UCDs in Virgo. The black line indicates the mass corresponding to the lower luminosity limit, while the red line indicates the mass corresponding to the lower radius limit. The two distributions were found to be consistent above a mass of 1~$\times$~10\textsuperscript{7}~\(\textup{M}_\odot\) (KS test, \hl{p~=~0.51}). The objects are divided into 7 bins from a minimum mass of 1.6~$\times$~10\textsuperscript{6}~\(\textup{M}_\odot\) to a maximum of 2~$\times$~10\textsuperscript{8}~\(\textup{M}_\odot\).}
    \label{fig:fullclustermasses}
\end{figure}

Fig.~\ref{fig:fullclustermasses} depicts the mass distribution of stripped nuclei and UCDs for the simulated cluster and Virgo. Above a mass of 1~$\times$~10\textsuperscript{7}~\(\textup{M}_\odot\) the simulated cluster contains a total of $51.8 \pm 7.7$ stripped nuclei, including $2.3 \pm 1.1$ above a mass of 1~$\times$~10\textsuperscript{8}~\(\textup{M}_\odot\). In the Virgo cluster, there are 40 UCDs above a mass of 1~$\times$~10\textsuperscript{7}~\(\textup{M}_\odot\), including 2 above a mass of 1~$\times$~10\textsuperscript{8}~\(\textup{M}_\odot\), indicating that stripped nuclei are consistent with making up all of the UCDs in Virgo that are above a mass of 1~$\times$~10\textsuperscript{7}~\(\textup{M}_\odot\), and all of the extremely massive M~>~1~$\times$~10\textsuperscript{8}~\(\textup{M}_\odot\) ones. Applying the Kolmogorov-Smirnov test to the mass distributions of UCDs and stripped nuclei above 1~$\times$~10\textsuperscript{7}~\(\textup{M}_\odot\) returns \hl{p~=~0.51}, indicating that the two distributions are consistent.

\subsubsection{Radial distributions around the central three galaxies}
\label{section:3raddists}

\begin{figure}
	\includegraphics[width=\columnwidth]{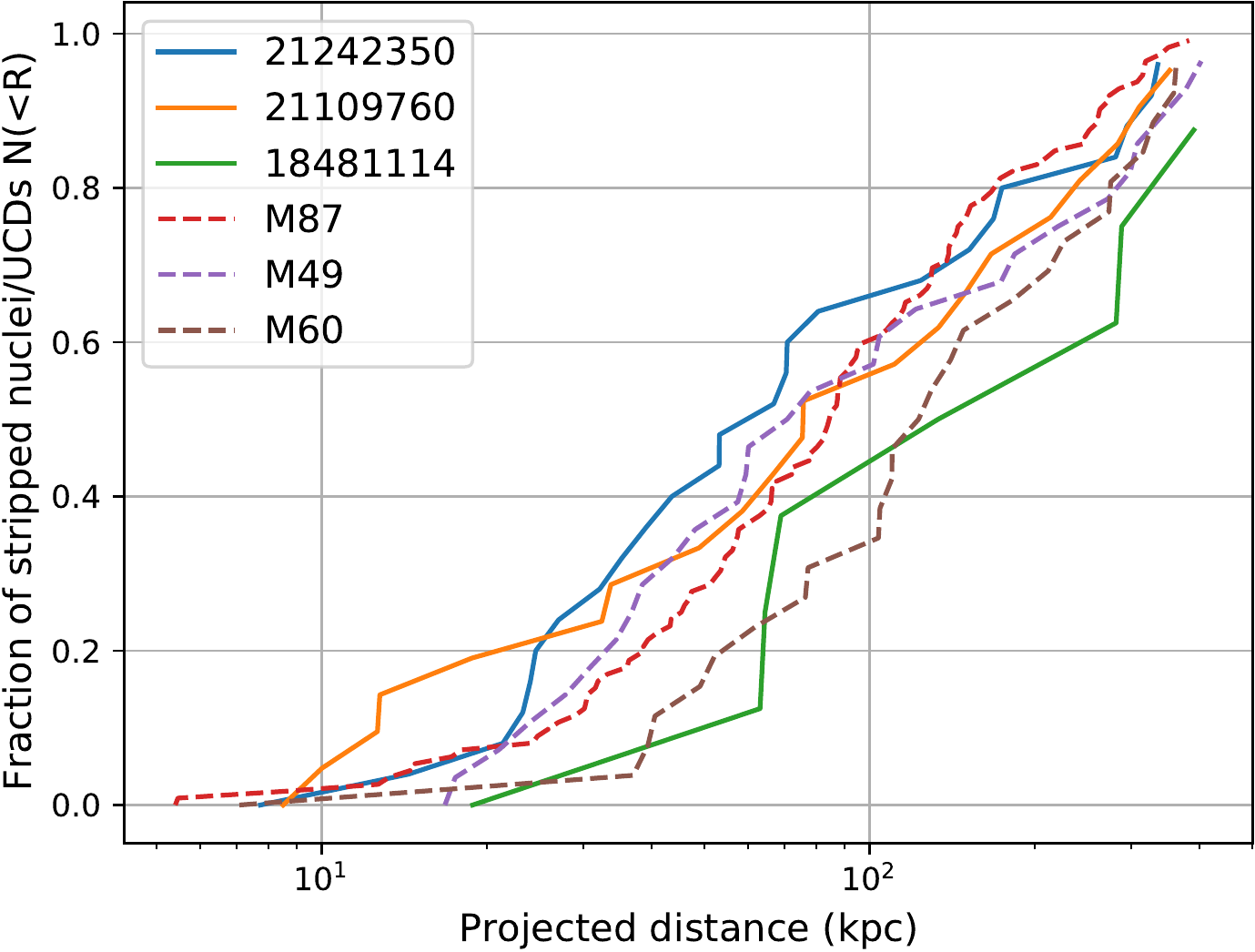}
    \caption{A combined plot of the radial distributions of aperture sampled M~>~1~$\times$~10\textsuperscript{7}~\(\textup{M}_\odot\) stripped nuclei around simulated galaxies and UCDs around the Virgo galaxies.}
    \label{fig:allchngzemergedradlimit}
\end{figure}

Fig.~\ref{fig:allchngzemergedradlimit} depicts the aperture \hl{sampled} distribution of M~>~1~$\times$~10\textsuperscript{7}~\(\textup{M}_\odot\) stripped nuclei and UCDs for the three most massive galaxies in the simulated cluster (Table.~\ref{tab:cluster000}) and in Virgo. Visually the three galaxies show similar distributions to the simulated galaxies. The radial distributions of the stripped nuclei surrounding the simulated galaxies were consistent between the three galaxies with p~>~0.05.

The radial distributions of the simulated stripped nuclei are consistent between the three different simulated galaxies. The radial distributions of UCDs around the observed galaxies are also consistent.

Comparing each of the radial distributions of the aperture sampled simulated M~>~1~$\times$~10\textsuperscript{7}~\(\textup{M}_\odot\) stripped nuclei to the Virgo galaxy UCDs allows for a total of 9 different distribution comparisons. We found that all comparisons returned p-values above 0.05 except for M60 and Galaxy ID = 21242350, which had p~=~0.046. M60 appears to lack many UCDs at a low galactocentric distance. This could be due to the difficulty of detecting UCDs at small radii around observed galaxies because of the high surface brightnesses at the centres of galaxies, or to the lower numbers of UCDs surrounding M60 making the radial distribution less reliable. Visually it appears most similar to the simulated galaxy Galaxy ID = 18481114, which is the least massive of the three simulated galaxies and contains the fewest stripped nuclei.

\subsection{Comparisons with other simulated clusters}
In addition to the massive cluster selected to compare with Virgo, we further analyze 6 other clusters with $\mathcal{M}$\textsubscript{200}~>~1~$\times$~10\textsuperscript{14}~\(\textup{M}_\odot\). 

\subsubsection{Numbers of stripped nuclei in clusters}
\begin{figure}
	\includegraphics[width=\columnwidth]{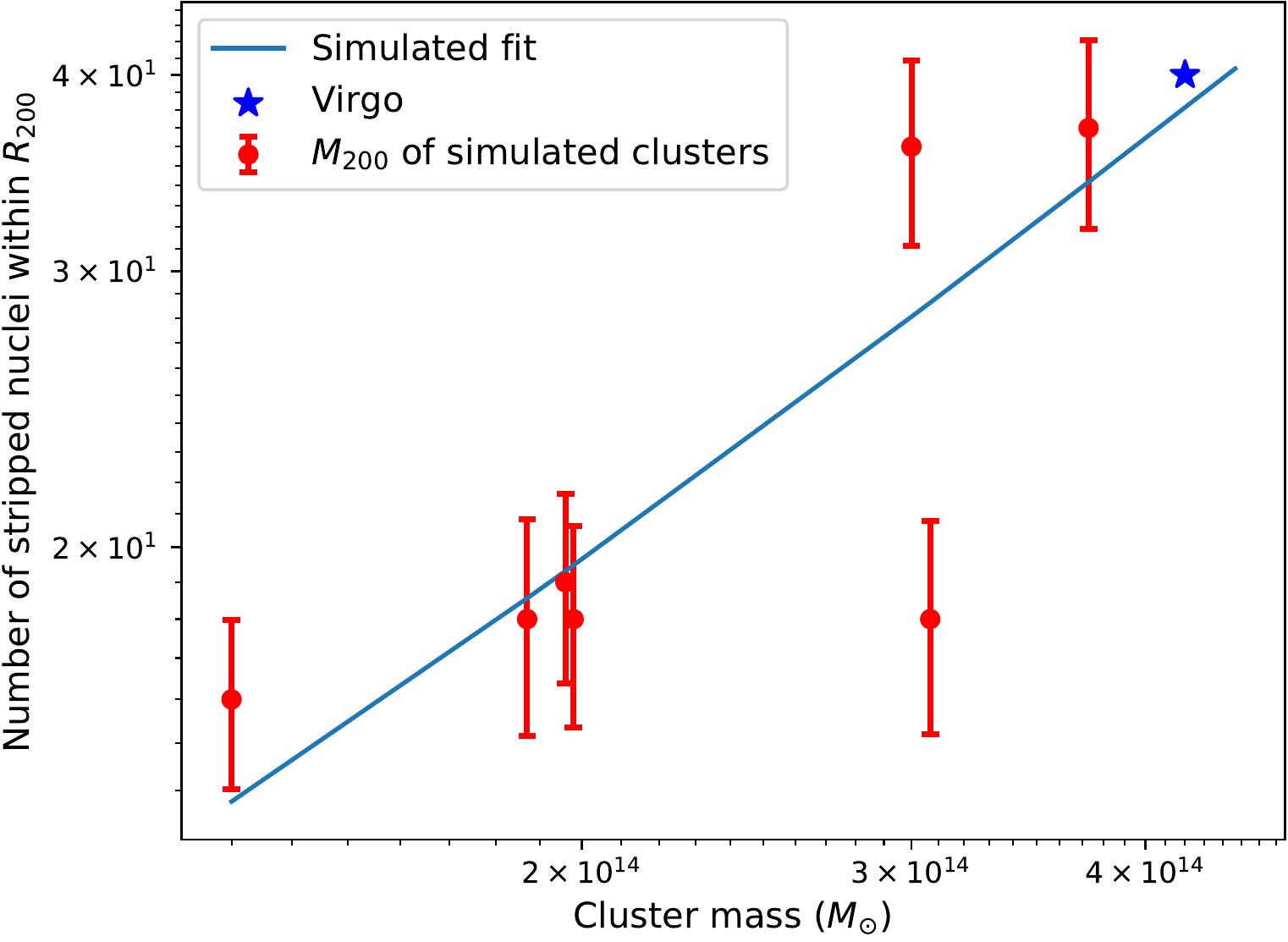}
    \caption{The number of M~>~1~$\times$~10\textsuperscript{7}~\(\textup{M}_\odot\) stripped nuclei compared with $\mathcal{M}$\textsubscript{200}, including the Virgo cluster}
    \label{fig:clustnummasseslobf}
\end{figure}

\begin{figure}
	\includegraphics[width=\columnwidth]{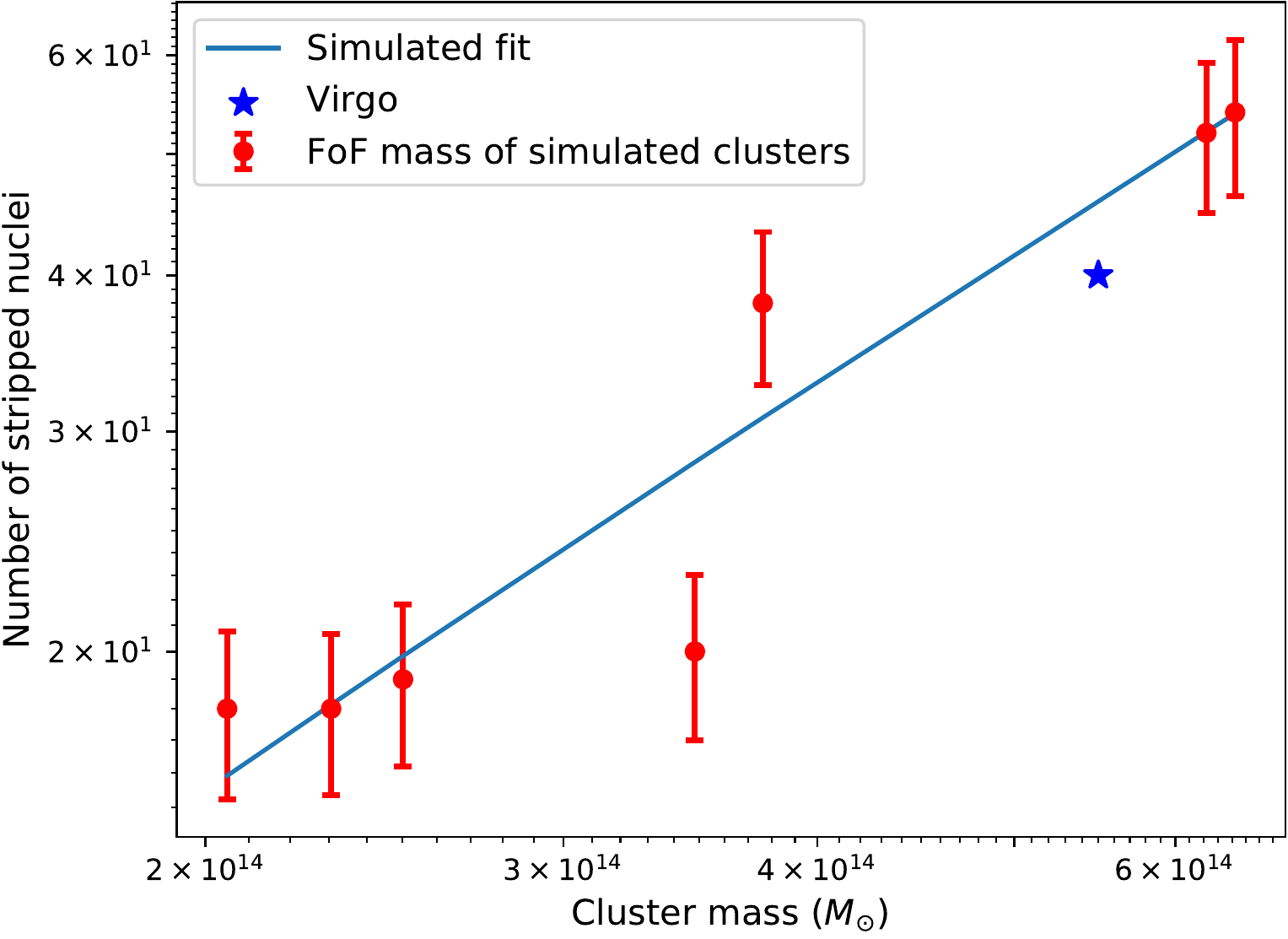}
    \caption{The number of M~>~1~$\times$~10\textsuperscript{7}~\(\textup{M}_\odot\) stripped nuclei compared with FoF mass, including the Virgo cluster}
    \label{fig:clustnummasseslobfv2}
\end{figure}

When comparing the number of stripped nuclei in simulated clusters to Virgo we consider both the FoF mass and $\mathcal{M}$\textsubscript{200}, listed in Table \ref{tab:clusters_analysed}. 
Fig.~\ref{fig:clustnummasseslobf} plots the number of M~>~1~$\times$~10\textsuperscript{7}~\(\textup{M}_\odot\) stripped nuclei as a function of $\mathcal{M}$\textsubscript{200}, while Fig.~\ref{fig:clustnummasseslobfv2} does the same for the FoF masses. \hl{The number of stripped nuclei for the FoF masses is taken within the whole cluster. For \mbox{$\mathcal{M}$\textsubscript{200}} the stripped nuclei sample is taken within \mbox{$\mathcal{R}$\textsubscript{200}}}. Vertical error bars are determined from the uncertainty in nucleus mass and nucleation fraction. The $\mathcal{M}$\textsubscript{200} plot has a linear relationship of $\log$\textsubscript{10} N =~($0.75 \pm 0.3$)~$\log$\textsubscript{10}~M\textsubscript{*}~$-(9 \pm 4)$, while the FoF plot has a relationship $\log$\textsubscript{10}~N~=~($1.0 \pm 0.2$)~$\log$\textsubscript{10}~M\textsubscript{*}~$-(14 \pm 2)$. The number-cluster mass relations are consistent with the number of observed UCDs in the Virgo Cluster (to 1 sigma), in both cases.  The overlapping relationships suggest that the distinction between FoF and $\mathcal{M}$\textsubscript{200} may not be strictly necessary when comparing numbers of stripped nuclei in galaxy clusters, although the predicted numbers for the FoF masses trend lower.


\subsubsection{Cumulative radial distributions around central galaxies}
\label{section:cenraddists}

To verify that the stripped nuclei distributions of simulated galaxies could explain the distributions of observed UCDs around observed galaxies we need to test the amount of scatter in the simulations. We do this by comparing radial distributions of simulated stripped nuclei surrounding massive galaxies in different simulated clusters.
We will primarily work with the aperture sample of M~>~1~$\times$~10\textsuperscript{7}~\(\textup{M}_\odot\) stripped nuclei, as the observed sample is largely complete in this mass range and may be incomplete for lower masses.

\begin{figure}
	\includegraphics[width=\columnwidth]{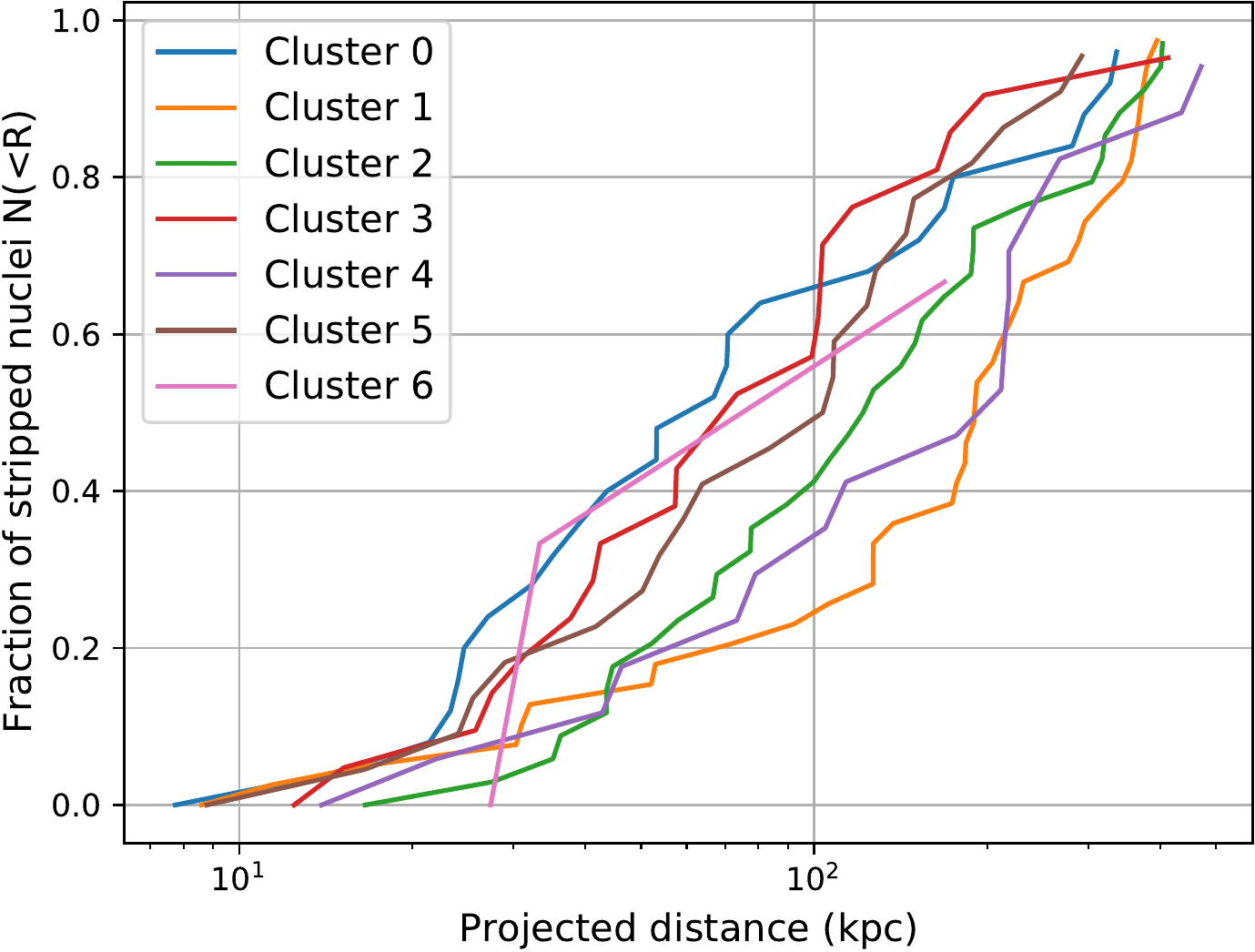}
    \caption{Aperture \hl{sampled} cumulative distribution of M~>~1~$\times$~10\textsuperscript{7}~\(\textup{M}_\odot\) stripped nuclei around the central most massive galaxies of the simulated clusters.}
    \label{fig:cumdistcnetralclusters}
\end{figure}

Fig.~\ref{fig:cumdistcnetralclusters} plots the aperture sampled cumulative radial distribution of M~>~1~$\times$~10\textsuperscript{7}~\(\textup{M}_\odot\) stripped nuclei around the most massive galaxy for the seven simulated clusters, shown in Table \ref{tab:clusters_analysed}.

\hl{From the 7 galaxies we can choose 21 different pairs of galaxy stripped nuclei distributions to compare}. Of these 17/21 were consistent above p~>~0.05 and 18/21 above p~>~0.01. Three of the comparisons returned p~<~0.01. All three of these p~<~0.01 pairs included a single galaxy, the massive one in Cluster 1.

\begin{figure}
	\includegraphics[width=\columnwidth]{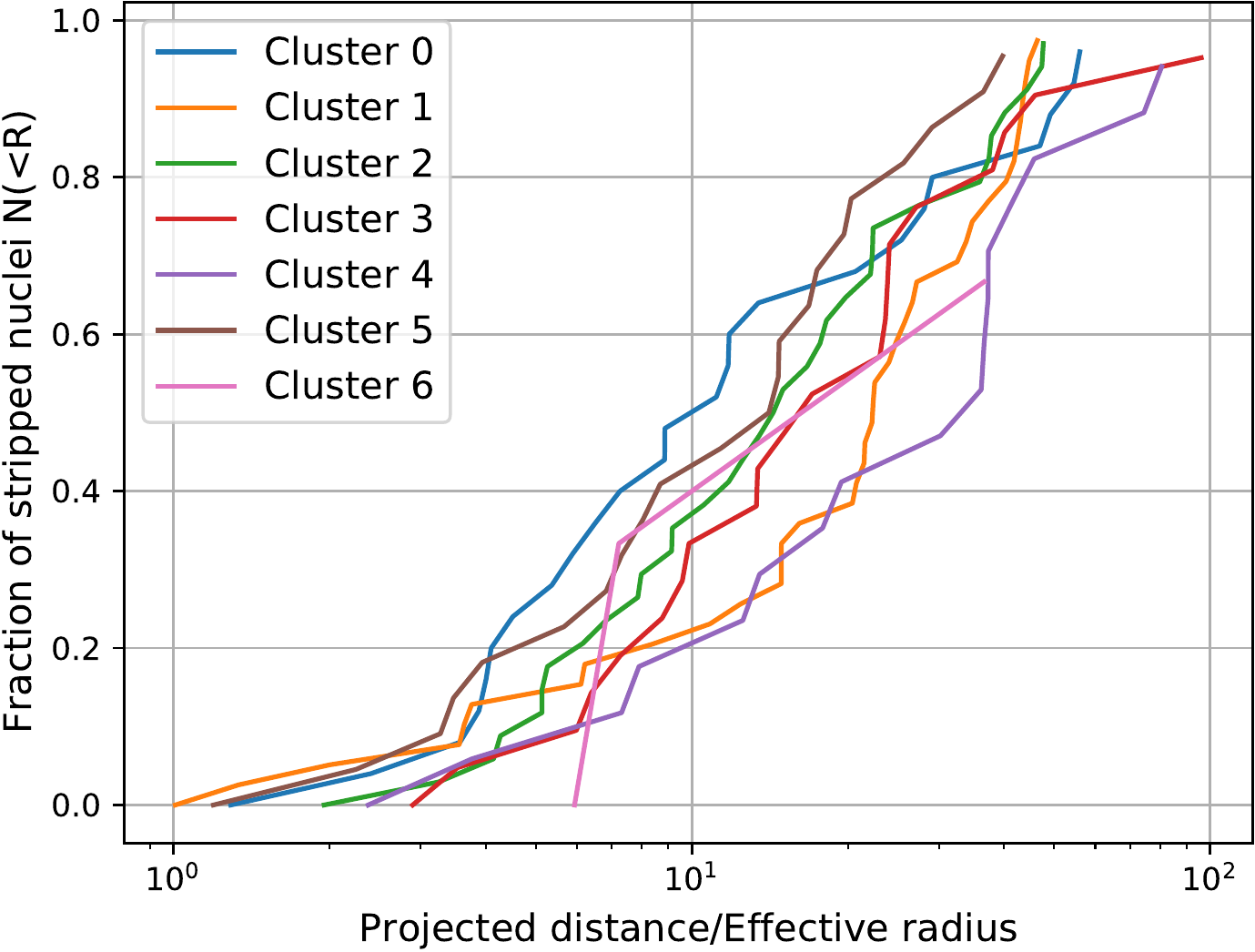}
    \caption{Aperture \hl{sampled} cumulative distribution of M~>~1~$\times$~10\textsuperscript{7}~\(\textup{M}_\odot\) stripped nuclei around the central most massive galaxies of the simulated clusters divided by the projected physical radius
    enclosing half of the stellar mass within 30 kpc.}
    \label{fig:multiclustercumdensityhalfmassradproj1e7}
\end{figure}

The central galaxies in these seven clusters had a range of properties such as radii, so we also considered the radial distributions when rescaled by the half mass radius of the stellar mass within 30 kpc, shown in Fig.~\ref{fig:multiclustercumdensityhalfmassradproj1e7}. In this case, 18/21 of the radial distributions were consistent above p~>~0.05 and 20/21 were consistent above p~>~0.01. Two of the p~<~0.05 pairs again included a single galaxy, the massive one in Cluster 1.

The merger tree distributions for M~>~1~$\times$~10\textsuperscript{7}~\(\textup{M}_\odot\) stripped nuclei were found not to require rescaling with all 21 being consistent above p~>~0.05.

\begin{figure}
	\includegraphics[width=\columnwidth]{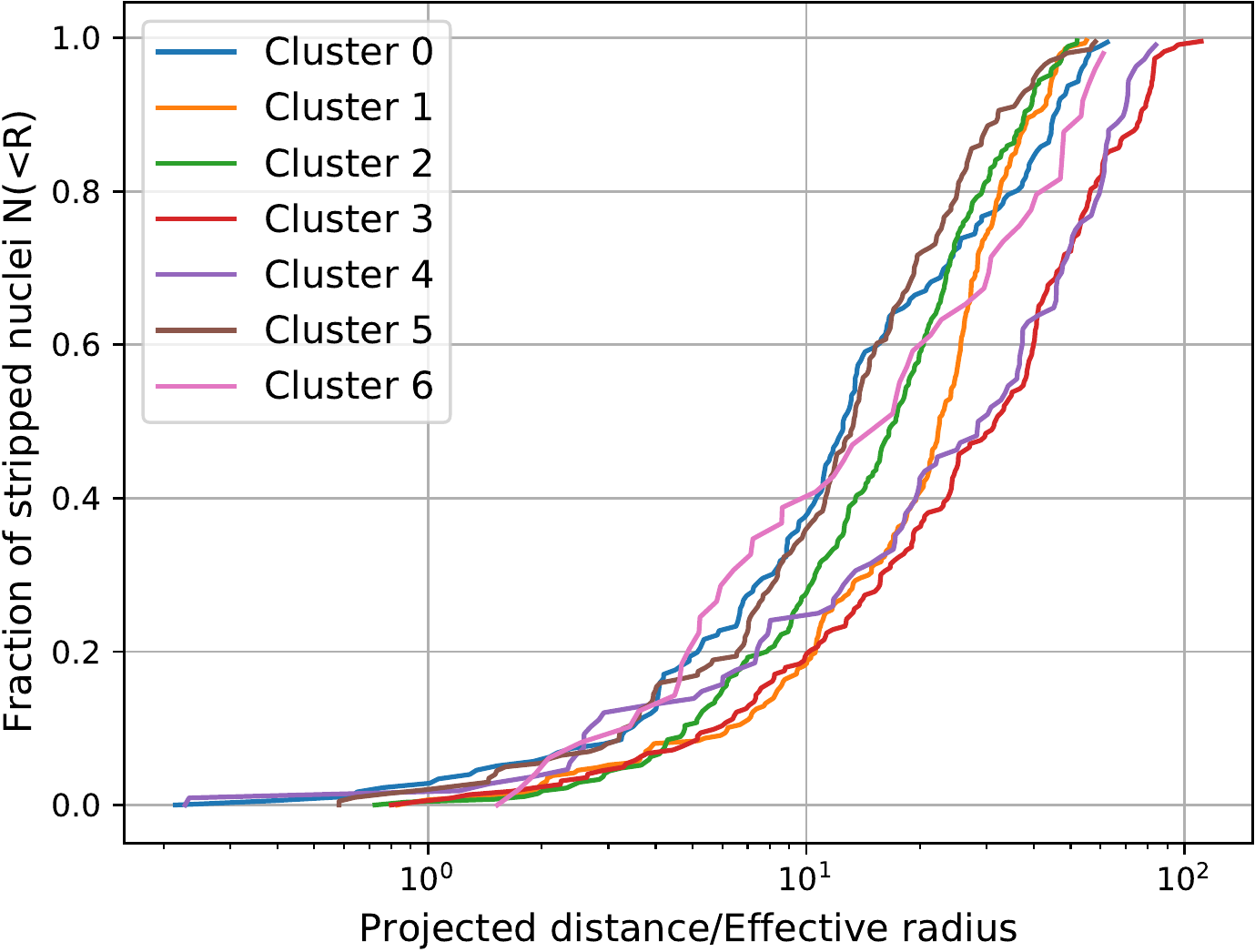}
    \caption{Aperture \hl{sampled} cumulative distribution of M~>~2~$\times$~10\textsuperscript{6}~\(\textup{M}_\odot\) stripped nuclei around the central most massive galaxies of the simulated clusters divided by the projected physical radius
    enclosing half of the stellar mass within 30 kpc.}
    \label{fig:multiclustercumdensityhalfmassradproj2E6}
\end{figure}

The lower mass distributions were less consistent than the higher mass ones. Few of the radial distributions for M~>~2~$\times$~10\textsuperscript{6}~\(\textup{M}_\odot\) stripped nuclei were consistent. This applied to both the merger tree and aperture \hl{sampled} distributions even when rescaled. Fig.~\ref{fig:multiclustercumdensityhalfmassradproj2E6} depicts the rescaled aperture \hl{sampled} distribution of M~>~2~$\times$~10\textsuperscript{6}~\(\textup{M}_\odot\) stripped nuclei.



\subsubsection{Numbers of stripped nuclei in individual galaxies against galaxy stellar mass}
\label{section:stellmass}

\begin{figure}
	\includegraphics[width=\columnwidth]{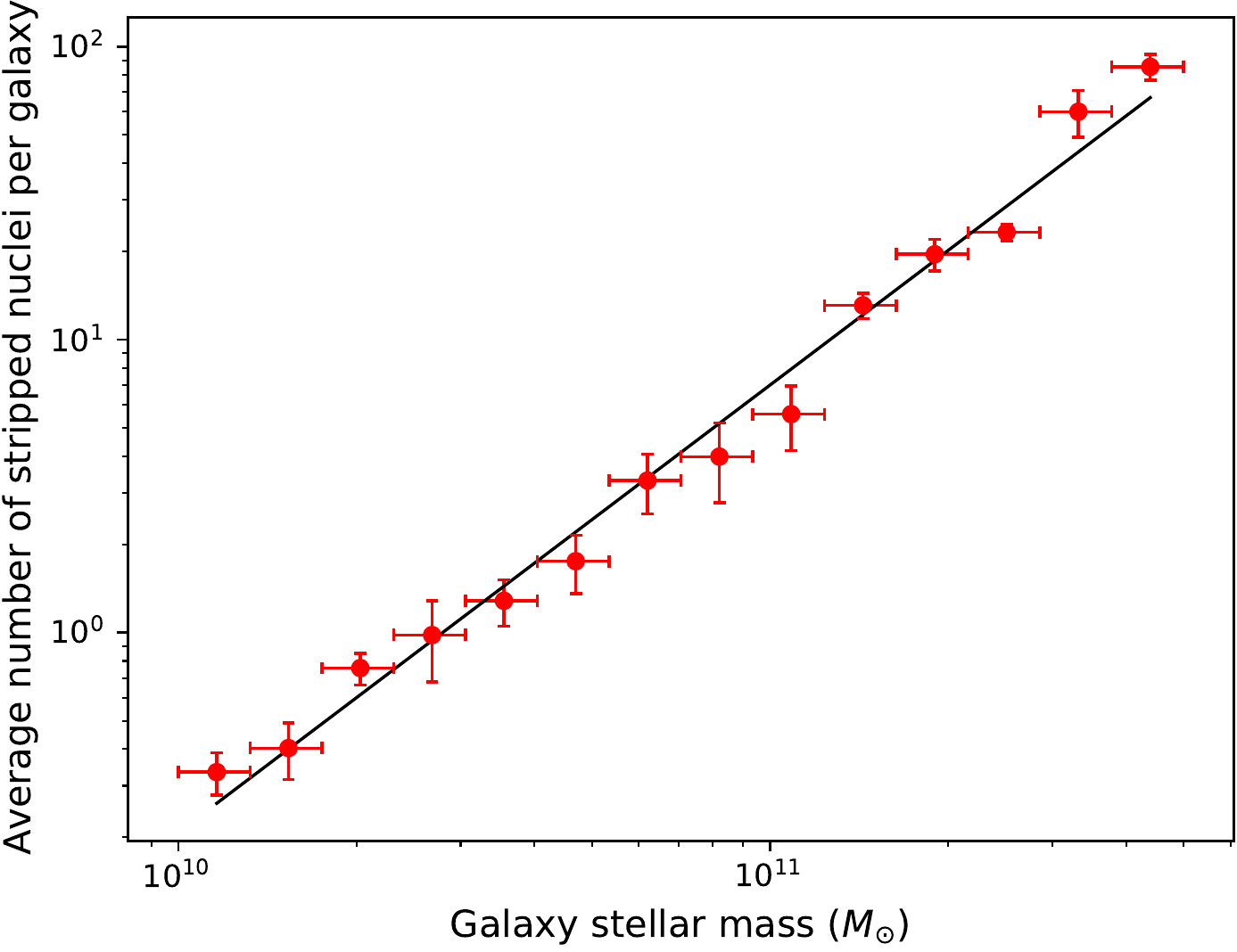}
    \caption{The total number of M~>~2~$\times$~10\textsuperscript{6}~\(\textup{M}_\odot\) stripped nuclei in individual simulated galaxies binned by galaxy stellar mass. The horizontal error bars are the mass range, the vertical error bars are the uncertainty of the mean.}
    \label{fig:nosnlinearfits}
\end{figure}

Fig.~\ref{fig:nosnlinearfits} plots the mean number of merger-tree M~>~2~$\times$~10\textsuperscript{6}~\(\textup{M}_\odot\) stripped nuclei in individual simulated galaxies for all seven clusters binned by galaxy stellar mass, with horizontal error bars from the binned mass range and vertical error bars from the uncertainty of the mean. A linear fit is applied to the total distribution, and found to have a relationship of $\log$\textsubscript{10}~N~=~($1.53 \pm 0.05$)~$\log$\textsubscript{10}~M\textsubscript{*}~$-(16.0 \pm 0.6)$. 

To determine whether there is a relationship between the stellar masses of the host clusters and the number of stripped nuclei, we then binned the numbers of stripped nuclei around galaxies in each of the seven clusters separately. We applied a linear fit to each of these seven plots using the same slope as the full fit to galaxies in all seven clusters. We next determined the y-intercept of these separate cluster fits and compared them to the y intercept of the fit to all seven clusters. For each separate fit, the y-intercept was found within 1 sigma of the y-intercept of the fit to all seven clusters, suggesting that the stripped nuclei number to stellar mass relation is not influenced by the mass of the cluster the galaxy is located within. This suggests that while Fig.~\ref{fig:clustnummasseslobf} and Fig.~\ref{fig:clustnummasseslobfv2} show a trend between the cluster mass and the number of stripped nuclei, the number of stripped nuclei in a cluster is more dependent on the numbers and masses of galaxies located within the cluster than the total cluster mass. More massive clusters have more stripped nuclei due to containing more massive galaxies and a greater number of galaxies, which will produce more stripped nuclei. The cluster mass itself does not have any influence on the stripping process. 


Fig.~\ref{fig:multimassiveprojectedwithvirgo} plots the binned number of M~>~1~$\times$~10\textsuperscript{7}~\(\textup{M}_\odot\) stripped nuclei around individual galaxies from all 7 clusters, using the aperture \hl{sample}, as well as the number of UCDs observed surrounding the three most massive galaxies in Virgo. \hl{In order to be consistent with the simulated galaxies, we take the stellar mass of the observed galaxies to be the enclosed stellar mass inside a radius of 30 kpc.} M87's stellar mass \hl{inside this
radius} is from \hl{figure 7 in} \citep{Gebhardt_2009}, M60's is from \hl{figure 10 in} \citep{Hwang_2008}, and M49's is from \hl{figure 17 in} \citep{Cote_2003}. \hl{The total masses of the galaxies within a radius of 100 kpc can be drawn from the same plots.} The horizontal error bars are from the binned mass range, while the vertical error bars are from the standard deviation of the binned data. A linear fit is applied to the total distribution and found to have a relationship of $\log$\textsubscript{10}~N~=~($1.2 \pm 0.3$)~$\log$\textsubscript{10}~M\textsubscript{*}~$-(12.0 \pm 4)$. 
Plotting the aperture \hl{sampled} numbers of stripped nuclei gives considerably more uncertainty than plotting the merger-tree numbers. This is because if two galaxies are lying close together, their distributions can overlap, resulting in stripped nuclei being double-counted for these two galaxies, inflating the overall numbers of stripped nuclei. To mitigate this issue, we implement the following criteria when plotting simulated galaxies:

\begin{enumerate}
  \item We included only galaxies with stellar mass M~>~1~$\times$~10\textsuperscript{11}~\(\textup{M}_\odot\), because smaller galaxies are more likely to overlap with larger ones
  \item The sampled radius of stripped nuclei was restricted to 200~kpc, to reduce the number of galaxies with overlapping stripped nuclei populations.
  \item When two galaxies would have overlapping stripped nuclei populations (distance between galaxies < 400~kpc) we excluded the less massive of the two galaxies from the plot.
\end{enumerate}

The plot shows substantially more scatter than the merger tree plot, largely due to the more variable numbers produced by the apertures. Of the Virgo galaxies, M49 and M60 \hl{are} within two sigma of the line of best fit, and M87 is within one sigma, indicating that the numbers of M~>~10\textsuperscript{7}~\(\textup{M}_\odot\) stripped nuclei in the simulated galaxies are consistent with the numbers of M~>~10\textsuperscript{7}~\(\textup{M}_\odot\) UCDs within the Virgo galaxies.

\subsubsection{Numbers of stripped nuclei in individual galaxies against galaxy halo mass}

In addition to comparing the number of stripped nuclei surrounding galaxies to galaxy stellar mass we compared the number to galaxy halo mass, as some studies have suggested that halo mass is a better predictor of UCD numbers than stellar mass \citep{Pfeffer2014,Liu2015}.

\begin{figure}
	\includegraphics[width=\columnwidth]{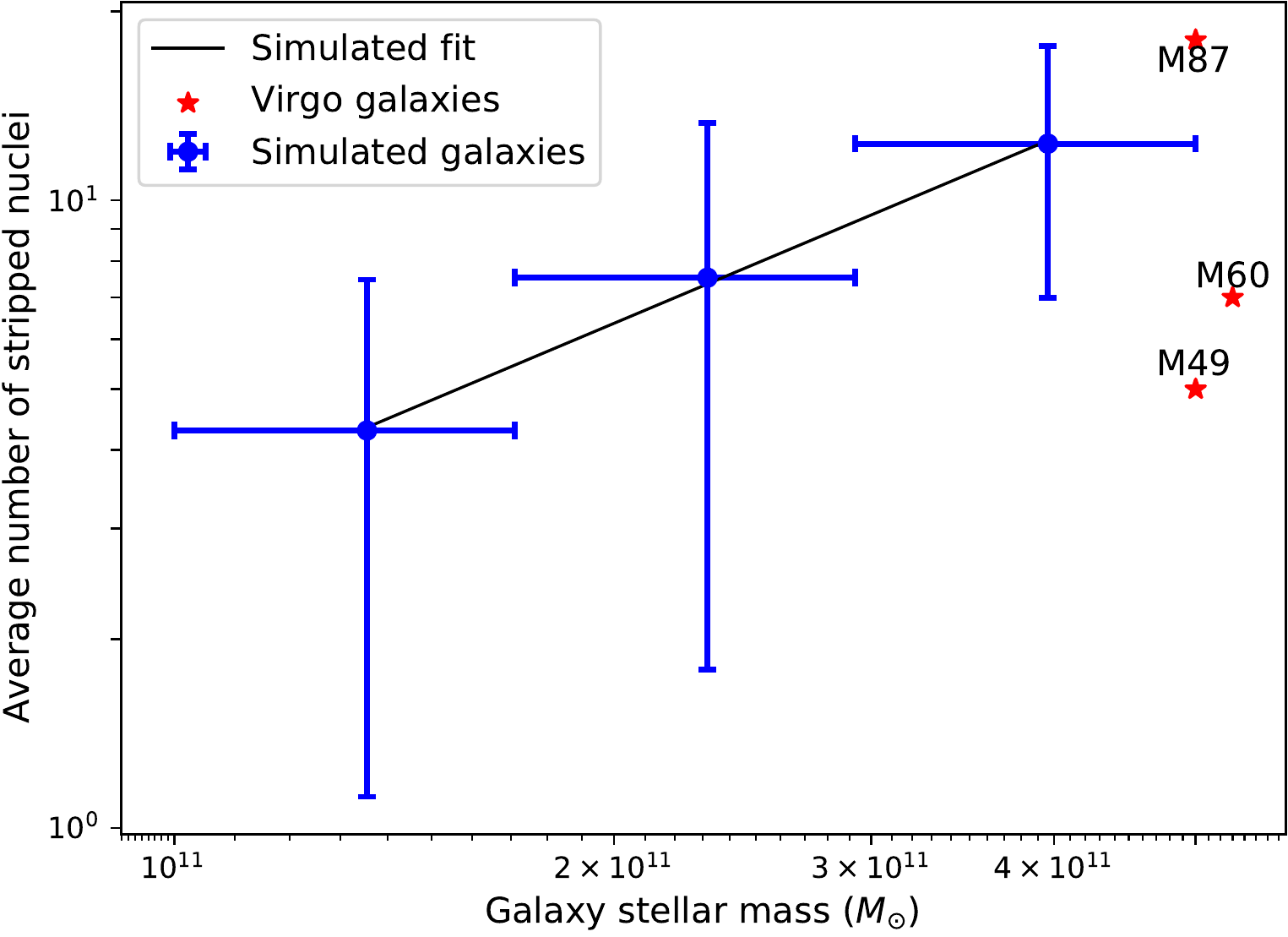}
    \caption{Number of aperture sampled M~>~1~$\times$~10\textsuperscript{7}~\(\textup{M}_\odot\) UCDs and stripped nuclei in individual galaxies for galaxies in our simulated cluster with stellar mass M~>~10\textsuperscript{11}~\(\textup{M}_\odot\) and the three Virgo cluster galaxies, against galaxy stellar mass. The simulated galaxy stripped nuclei numbers are taken from sampling stripped nuclei in the area within 200~kpc radii of them. M87's mass is from \citep{Gebhardt_2009}, M60's from \citep{Hwang_2008} and M49's from \citep{Cote_2003}. The horizontal error bars are the mass range, the vertical error bars are the standard deviation of each bin.}
    \label{fig:multimassiveprojectedwithvirgo}
\end{figure}

\begin{figure}
	\includegraphics[width=\columnwidth]{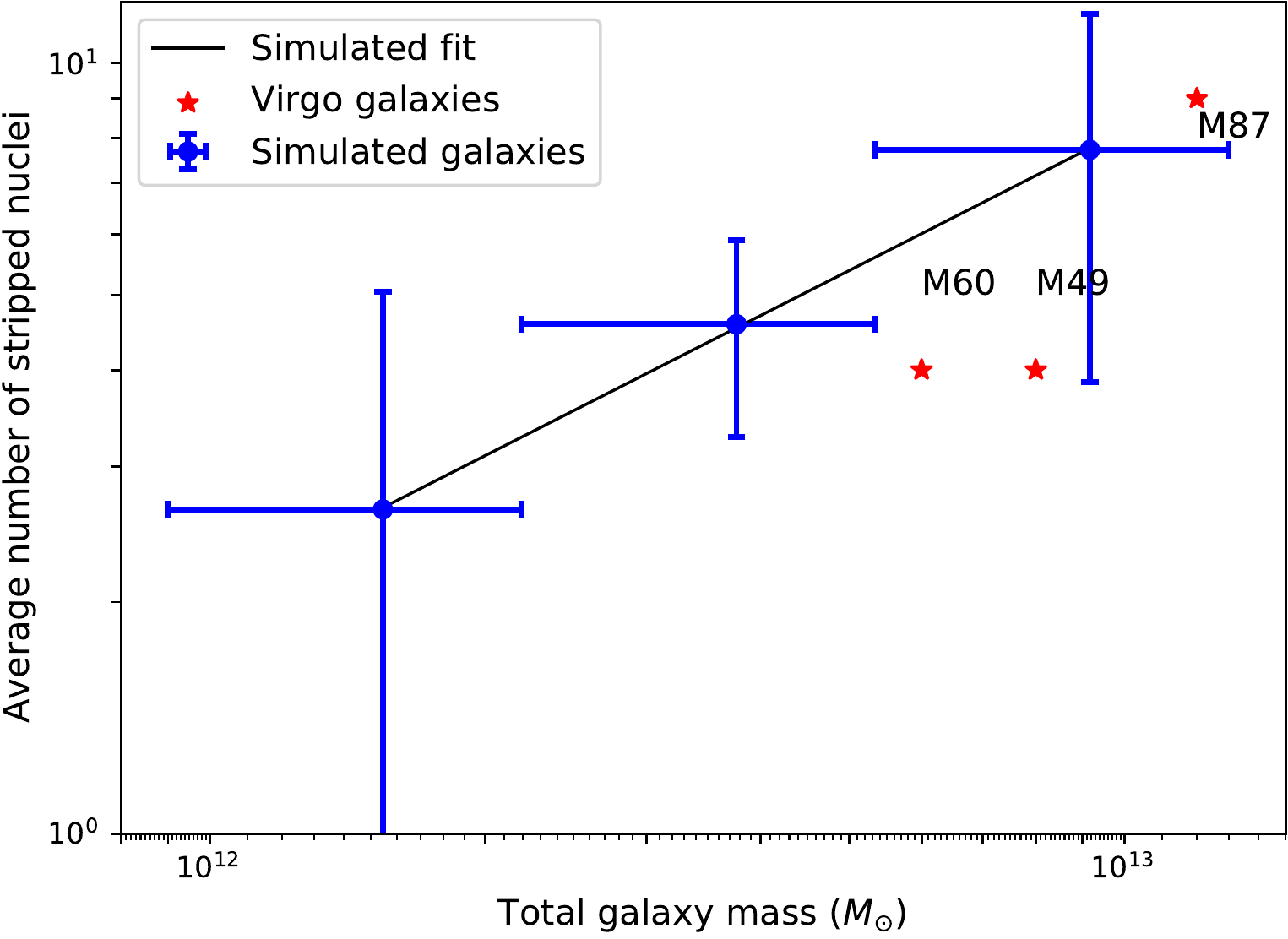}
    \caption{Same as Fig.~\ref{fig:multimassiveprojectedwithvirgo}, but with the total galaxy mass within an aperture of 100~kpc against the number of aperture sampled M~>~1~$\times$~10\textsuperscript{7}~\(\textup{M}_\odot\) UCDs and stripped nuclei within a 100~kpc radius.}
    \label{fig:multimassiveprojectedwithvirgov2}
\end{figure}

Fig.~\ref{fig:multimassiveprojectedwithvirgov2} plots the binned aperture \hl{sampled} number of M~>~1~$\times$~10\textsuperscript{7}~\(\textup{M}_\odot\) stripped nuclei around individual galaxies from all 7 clusters, as well as the number of UCDs observed surrounding the three most massive galaxies in Virgo, against the total galaxy mass within a 100~kpc aperture. The stripped nuclei are selected using the same criteria as the stellar mass sample, however, we now sample stripped nuclei within a 100 kpc radius to match the halo radius. \hl{We take the total masses of the three observed Virgo galaxies from the same sources as the stellar masses, as described in section \mbox{\ref{section:stellmass}}}. The horizontal error bars are from the binned mass range, while the vertical error bars are from the standard deviation of the binned data. A linear fit is applied to the total distribution, and was found to have a relationship of $\log$\textsubscript{10}~N~=~($0.6 \pm 0.03$)~$\log$\textsubscript{10}~M\textsubscript{*}~$-(7.4 \pm 0.3)$. The plot shows a similar amount of scatter to Fig.~\ref{fig:multimassiveprojectedwithvirgo}, but the Virgo galaxies are more consistent with this fit than they are with the stellar mass relation, with all three found within one sigma of the line of best fit.

\subsection{Low-mass stripped nuclei}
We also find that a number of low mass M~<~2~$\times$~10\textsuperscript{6}~\(\textup{M}_\odot\) stripped nuclei are created by tidal stripping. These stripped nuclei may not be identified in observations as UCDs due to their low masses but instead, be observed as globular clusters. Around the three most massive galaxies in the most massive cluster, we find $321 \pm 43$, $289 \pm 38$ and $163 \pm 21$ low mass stripped nuclei respectively. This is low in comparison to the number of globular clusters hosted in galaxies of this mass, with M87 predicted to host as many as 18000 globular clusters \citep{Oldham2016} and M60 to host a globular cluster population of approximately 3700 \citep{Forbes2004}. However, it should be noted that due to the limitation of this study to simulated galaxies with stellar mass M~>~1~$\times$~10\textsuperscript{7}~\(\textup{M}_\odot\) we likely underestimate the number of low mass stripped nuclei and instead the number should be taken as a lower limit.

\begin{figure}
	\includegraphics[width=\columnwidth]{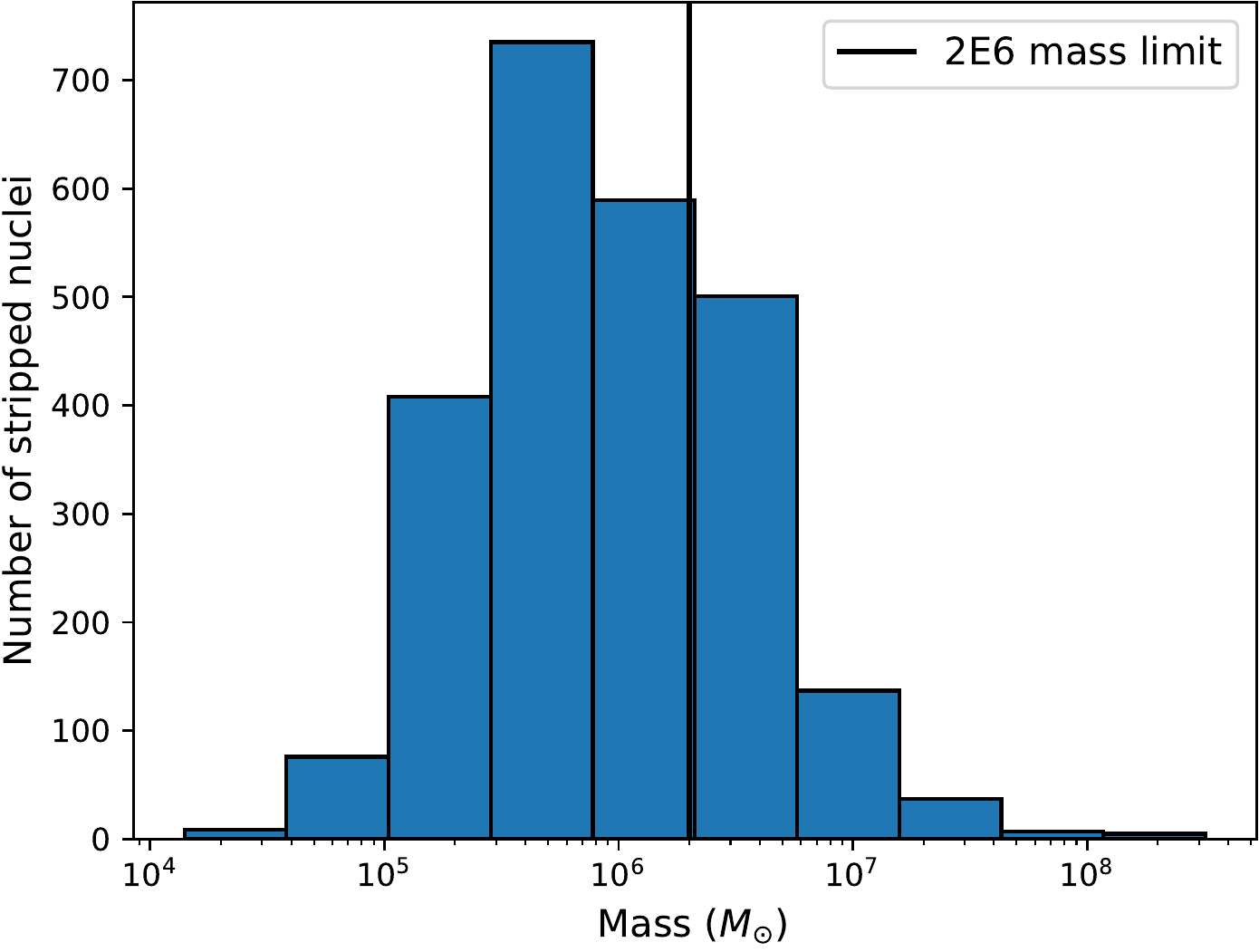}
    \caption{Mass  distribution  for  all stripped nuclei from the most massive simulated cluster including low mass  M~<~2~$\times$~10\textsuperscript{6}~\(\textup{M}_\odot\) stripped  nuclei.}
    \label{fig:lowmasssn}
\end{figure}

Fig.~\ref{fig:lowmasssn} plots the mass distribution for low mass stripped nuclei in the most massive simulated cluster. While the distribution appears to peak around M~$\sim$~5~$\times$~10\textsuperscript{5}~\(\textup{M}_\odot\) this is an artificial result of the 1~$\times$~10\textsuperscript{7}~\(\textup{M}_\odot\) stellar mass limit on progenitor galaxies, and the true peak is likely lower. We find a total of $1780 \pm 240$ low mass stripped nuclei in this cluster. Of these $1400 \pm 190$ are in the 10\textsuperscript{4}~\(\textup{M}_\odot\) < M < 10\textsuperscript{6}~\(\textup{M}_\odot\) range typically defined for globular clusters. Virgo is known to host $67300 \pm 1440$ globular clusters \citep{Durrell2014} indicating that several per cent of the globular clusters in Virgo sized galaxy clusters are likely stripped nuclei.  

Depending on their mass and size, these low-mass stripped nuclei will either be seen as massive globular clusters or as low mass stripped nuclei. It should be noted that while M~$\sim$~2~$\times$~10\textsuperscript{6}~\(\textup{M}_\odot\) is often considered the dividing mass between globular clusters and UCDs \citep[e.g.][]{has2005} the UCD/GC divide is largely arbitrary and differs from study to study, meaning that whether these objects would be classified as UCDs or globular clusters would depend on other factors such as their size. \citet{Forbes2013} for example, finds several low luminosity UCDs with sizes > 7~pc and M\textsubscript{V}~$\sim$ -8 to -9 (M~$\sim$~few~*~10\textsuperscript{5}~\(\textup{M}_\odot\)), that would be similar to these low mass stripped nuclei.
\hl{Although the number of stripped nuclei contributing to the overall globular cluster population will be small, each of these merging galaxies likely brought in their own population of globular clusters. Modern cosmological hydrodynamical galaxy formation simulations, including EAGLE, consistently find that as much as 80 per cent of the mass of massive galaxies is accreted} \citep[e.g.][]{Oser2010,Rodriguez_Gomez_2017,Clauwens_2018,Davison2020}. \hl{Therefore a large portion of globular clusters found in massive galaxies will likely have been contributed by the progenitor galaxies of the existing stripped nuclei population.}

\subsection{Recently formed stripped nuclei}
A small number of stripped nuclei will have been accreted within the past two billion years. Instead of being seen as fully-formed objects, these stripped nuclei will likely appear as transitional objects undergoing stripping, or with significant amounts of galaxy debris surrounding them. For the most massive galaxies, we find 1-3 objects of this sort per galaxy, and in the most massive cluster, we find $6.3 \pm 3.2$ stripped nuclei that have merged within two billion years. \hl{Depending on the pericenter of the stripped nuclei orbits, objects older than 2 Gyr may also retain galaxy debris.}

Additionally, since our method focuses on progenitor galaxies found in merger-trees, we will likely miss a small number of stripped nuclei in the process of forming that are made up of merging galaxies that have suffered significant tidal stripping but are still considered by EAGLE to be separate galaxies. Transitional objects of this sort have been observed in the nearby universe \citep{Galianni2010, Jennings2015}.

Depending on the mass and surface density of the stellar debris of our transitional objects they may or may not be observable, i.e. a 10\textsuperscript{10}~\(\textup{M}_\odot\) galaxy may be very clear, but  a $\sim$10\textsuperscript{7}~\(\textup{M}_\odot\) galaxy may not be detectable without extremely deep imaging.
\section{Discussion}

\subsection{Can the radial distributions of UCDs in massive galaxy clusters be explained by stripped galaxy nuclei?}
\subsubsection{Using the simulations to test the consistency of stripped nuclei distributions}
In Section \ref{section:cenraddists}, we compared radial distributions of central galaxies to determine the scatter in the simulations. We found that the preferred method of comparing the radial distributions of stripped nuclei around different simulated galaxies included only M~>~1~$\times$~10\textsuperscript{7}~\(\textup{M}_\odot\) stripped nuclei and rescaled these distributions by the effective radius, as shown in Fig.~\ref{fig:multiclustercumdensityhalfmassradproj1e7}. In contrast to the higher mass sample, the radial distributions of M~>~2~$\times$~10\textsuperscript{6}~\(\textup{M}_\odot\) stripped nuclei showed large amounts of scatter. 

The reason why some of the M~>~1~$\times$~10\textsuperscript{7}~\(\textup{M}_\odot\) stripped nuclei distributions and few of the lower mass M~>~2~$\times$~10\textsuperscript{6}~\(\textup{M}_\odot\) stripped nuclei distributions were inconsistent before rescaling is unknown. However one likely explanation is that the massive galaxies in each of the clusters have different structures or cluster environments, which would impact the distributions of stripped nuclei and the way progenitor galaxies merged. This would have a greater impact on the distributions of low mass stripped nuclei, and the lower mass merging galaxies which produced those nuclei because these objects would be more sensitive to changes in environment. Supporting this hypothesis is the fact that all M~>~1~$\times$~10\textsuperscript{7}~\(\textup{M}_\odot\) merger tree stripped nuclei distributions were consistent even before rescaling, as this sample does not include stripped nuclei from satellite galaxies and is, therefore, less influenced by the environment surrounding the galaxy than the aperture sample. 

It should also be noted that 3/21 of the aperture sampled M~>~1~$\times$~10\textsuperscript{7}~\(\textup{M}_\odot\) radial distributions that were inconsistent before rescaling all included the same galaxy, the most massive one in cluster 1, which then became more consistent with the other galaxies when rescaled. This galaxy's stripped nuclei population appeared more shallow and extended than the other 5 in Fig.~\ref{fig:cumdistcnetralclusters}, which is the reason why it was inconsistent before rescaling. The radial distributions of the other galaxies did not require rescaling to be consistent.

A caveat here is that we are comparing massive galaxies found largely at centres of potential in their respective clusters. Less massive or more isolated galaxies could have more or less consistent distributions. Considering galaxies less massive than those found here could be difficult, however, as lower mass galaxies contain very few M~>~1~$\times$~10\textsuperscript{7}~\(\textup{M}_\odot\) stripped nuclei, as we found for the most massive galaxy in the least massive cluster. Comparing the distributions of M~>~2~$\times$~10\textsuperscript{6}~\(\textup{M}_\odot\) stripped nuclei around less massive galaxies could be valuable, however, especially if those galaxies are more isolated and thus could be used to test whether lower mass stripped nuclei are more influenced by the galaxy environment.

An additional caveat is that rescaling massive galaxies by the half stellar-mass radius within 30 kpc may not fully take into account variables in the environment around those galaxies that will impact the distributions of stripped nuclei. For example, properties such as the presence or lack of nearby massive galaxies, or the structure of the dark matter halo. When comparing distributions of different galaxies, a more detailed survey of the environment surrounding those galaxies may be important to determine if the environments could impact the consistency of the distributions.

From this we conclude that it is likely that lower mass  M~<~1~$\times$~10\textsuperscript{7}~\(\textup{M}_\odot\) stripped nuclei are more influenced by galaxy structure and environment than higher mass M~>~1~$\times$~10\textsuperscript{7}~\(\textup{M}_\odot\) stripped nuclei, and this is likely to impact comparisons of radial distributions of stripped nuclei and UCDs between galaxies. Higher mass M~>~1~$\times$~10\textsuperscript{7}~\(\textup{M}_\odot\) stripped nuclei are a more viable option when comparing radial distributions between simulations and observations, and rescaling the distributions is a viable method of increasing consistency between the radial distributions of differently structured massive galaxies for M~>~1~$\times$~10\textsuperscript{7}~\(\textup{M}_\odot\) stripped nuclei. It may be desirable to compare the radial distributions of UCDs around central galaxies found in different clusters such as Virgo and Fornax to see if the distributions of observed UCDs are consistent. Additionally, comparisons of the radial distributions for UCDs of lower masses (M~<~1~$\times$~10\textsuperscript{7}~\(\textup{M}_\odot\)) and those of higher masses (M~>~1~$\times$~10\textsuperscript{7}~\(\textup{M}_\odot\)) between galaxies could be made to see if lower mass UCD distributions are observed to be more disrupted.

\subsubsection{Radial distributions of simulated stripped nuclei and observed UCDs}
Recent observations suggest that in the Virgo cluster UCDs are found clustered around the most massive galaxies rather than being spread evenly throughout the cluster \citet{liu2020generation}. Our simulation results agree with this, with Fig.~\ref{fig:10galsproj} showing that stripped nuclei are associated with massive galaxies, which suggests that stripped nuclei can account for the overall distributions of UCDs throughout the Virgo cluster.
Due to the strong evidence of clustering shown by UCDs around massive galaxies in clusters, it is more valuable to compare distributions of stripped nuclei around single galaxies, rather than analysing the distribution of stripped nuclei throughout clusters. 

We found that in addition to the observed similar clustering between UCDs and stripped nuclei the radial distribution of UCDs around M87 can be explained by the distribution of stripped nuclei around the central galaxy of the most massive galaxy cluster analysed, as shown in Fig.~\ref{fig:distnew}, with these two distributions being consistent to p~=~0.460 for M~>~1~$\times$~10\textsuperscript{7}~\(\textup{M}_\odot\) stripped nuclei. In Fig.~\ref{fig:allchngzemergedradlimit} we extended this comparison to the three massive simulated galaxies in this massive cluster compared against the radial distributions of the three Virgo cluster galaxies M87, M60 and M49 giving a total of 9 different distribution comparisons. All but one of the KS tests between the simulated galaxies and Virgo galaxies returned p~>~0.05, and the remainder returned p~=~0.046, which indicates the distribution of UCDs around central galaxies in the Virgo cluster are consistent with being stripped nuclei.

\citet{Thomas2008} found that the static model of galaxy stripping under-predicts the number of stripped nuclei at large radii, however, they noted that the static model has several challenges that could cause the number of stripped nuclei at large radii to be underestimated. \citet{Pfeffer2016} subsequently found that the radial distribution of GCs+UCDs and stripped nuclei were very consistent within 83 kpc for masses~M~>~5~$\times$~10\textsuperscript{6}~\(\textup{M}_\odot\), but within 300~kpc and for masses M~>~1~$\times$~10\textsuperscript{7}~\(\textup{M}_\odot\) stripped nuclei had a significantly more extended radial distribution than that of UCDs. In contrast to the previous studies we found that the radial distributions of stripped nuclei and UCDs were consistent up to 400 kpc for stripped nuclei with masses M~>~1~$\times$~10\textsuperscript{7}~\(\textup{M}_\odot\), so we conclude that the radial distributions of UCDs in the Virgo cluster are consistent with the radial distributions produced by stripped nuclei. The reason for the different results compared with \citet{Pfeffer2016} is likely due to that work using dark matter only simulations that cannot include important processes, such as the lack of a stellar component causing dark matter halos to disrupt at larger distances and a longer merging timescale allowing more time for dynamical friction to act.

\subsection{Can the numbers and masses of UCDs in large galaxy clusters be explained by stripped galaxy nuclei?}

\subsubsection{Numbers of stripped nuclei with masses M~>~1~$\times$~10\textsuperscript{7}~\(\textup{M}_\odot\)}
Previous studies have suggested that stripped nuclei make up approximately two-thirds of M~>~1~$\times$~10\textsuperscript{7}~\(\textup{M}_\odot\) UCDs \citep[e.g.][]{Mieske2012, Pfeffer2014, Pfeffer2016}. Our findings are consistent with these results, but also suggest that stripped nuclei could make up the entirety of M~>~1~$\times$~10\textsuperscript{7}~\(\textup{M}_\odot\) UCDs.  

Above 1~$\times$~10\textsuperscript{7}~\(\textup{M}_\odot\) the most massive simulated galaxy in the most massive cluster contains $14 \pm 4$ stripped nuclei, while M87 contains 18, indicating that stripped nuclei are consistent with making up all of the 1~$\times$~10\textsuperscript{7}~\(\textup{M}_\odot\) UCDs that surround M87. Fig.~\ref{fig:multimassiveprojectedwithvirgo} plots the binned average number of M~>~1~$\times$~10\textsuperscript{7}~\(\textup{M}_\odot\) stripped nuclei found within massive simulated galaxies, along with the three Virgo galaxies against stellar mass. All three Virgo galaxies are found within one or two sigma of the line of best fit indicating that stripped nuclei can completely explain the number of M~>~1~$\times$ 10\textsuperscript{7}~\(\textup{M}_\odot\) UCDs within these galaxies. In particular, the line of best fit predicts that M87 should have a total of $14.1 \pm 5.3$ UCDs which fits well with the actual number of 18.

\citet{Liu2015} found that the number of UCDs associated with observed galaxies is more closely correlated with the total mass of the host system than the stellar mass. \citet{Pfeffer2014} found that the number of stripped nuclei formed around simulated galaxies was also more correlated with halo mass than stellar mass. In Fig.~\ref{fig:multimassiveprojectedwithvirgov2} we plotted the binned average number of M~>~1~$\times$~10\textsuperscript{7}~\(\textup{M}_\odot\) stripped nuclei against total mass within 100~kpc. The plot shows similar amounts of scatter to the stellar mass plot making it difficult to confirm whether the number of stripped nuclei is more closely correlated with halo mass than stellar mass, however, this plot was more consistent with the observed galaxies than the stellar mass plot. All three Virgo galaxies are within one sigma of the line of best fit, a higher consistency than found with the stellar masses.

The most massive simulated cluster contained a total of $51.8 \pm 7.7$ stripped nuclei, while in the Virgo cluster there are 40 UCDs above a mass of 1~$\times$~10\textsuperscript{7}~\(\textup{M}_\odot\). Fig.~\ref{fig:clustnummasseslobf} and Fig.~\ref{fig:clustnummasseslobfv2} plot the number of stripped nuclei in 7 simulated clusters against the number of UCDs in Virgo for both the FoF mass and $\mathcal{M}$\textsubscript{200}. In both cases Virgo is found within one sigma of the line of best fit, indicating that stripped nuclei are consistent with making up the entirety of M~>~1~$\times$~10\textsuperscript{7}~\(\textup{M}_\odot\) UCDs in the Virgo cluster.

In conclusion, our model of stripped nuclei formation correctly predicts M~>~1~$\times$~10\textsuperscript{7}~\(\textup{M}_\odot\) UCD numbers for both individual galaxies and in the full Virgo cluster.

\subsubsection{Mass distributions of M~>~1~$\times$~10\textsuperscript{7}~\(\textup{M}_\odot\) stripped nuclei}
Fig.~\ref{fig:212masses} plots the binned number of stripped nuclei in the most massive galaxy in the most massive simulated cluster along with UCDs surrounding M87, the central galaxy in the Virgo cluster. A KS test found that the mass distributions of stripped nuclei and UCDs are consistent above M~>~1~$\times$~10\textsuperscript{7}~\(\textup{M}_\odot\) to p = 0.97. Fig.~\ref{fig:fullclustermasses} extends this test to the full cluster and Virgo, and finds that the mass distributions are consistent to p = 0.57. This indicates that the mass distributions of simulated stripped nuclei and UCDs are consistent above M~>~1~$\times$~10\textsuperscript{7}~\(\textup{M}_\odot\).

\subsubsection{Stripped nuclei in the mass range 2~$\times$~10\textsuperscript{6}~\(\textup{M}_\odot\) < M < 1~$\times$~10\textsuperscript{7}~\(\textup{M}_\odot\)}
Because of observational limitations on the detection of UCDs in the Virgo cluster, we can make no solid conclusions about what percentage of the M~<~1~$\times$~10\textsuperscript{7}~\(\textup{M}_\odot\) UCDs in Virgo are stripped nuclei. However, in our most massive cluster we find that $349 \pm 46$ of the $400 \pm 53$ M~>~2~$\times$~10\textsuperscript{6}~\(\textup{M}_\odot\) stripped nuclei have masses in the 2~$\times$~10\textsuperscript{6}~\(\textup{M}_\odot\) < M < 1~$\times$~10\textsuperscript{7}~\(\textup{M}_\odot\) range, indicating that 2~$\times$~10\textsuperscript{6}~\(\textup{M}_\odot\) < M < 1~$\times$~10\textsuperscript{7}~\(\textup{M}_\odot\) stripped nuclei make up approximately $87 \pm 20$ per cent of the stripped nuclei population in this cluster above a mass of 2~$\times$~10\textsuperscript{6}~\(\textup{M}_\odot\). \citet{Pfeffer2014} predicted that stripped nuclei could only make up 5~-~12 per cent of UCDs in the Fornax cluster above a mass of M > 2~$\times$~10\textsuperscript{6}~\(\textup{M}_\odot\). For this to be true for the Virgo cluster Virgo would have to host a population of 3300~-~8000 UCDs, substantially more than the current known 243. Therefore we predict that there either exists a large population of undiscovered low mass UCDs or a much larger percentage of the low mass UCD population is made up of stripped nuclei than previously predicted.

\subsubsection{Stripped nuclei with masses M < 2 $\times$ 10\textsuperscript{6} \(\textup{M}_\odot\)}
A number of globular clusters have been speculated to be stripped nuclei, including Omega Centauri \citep{Lee1999, hilker2000}, M22 \citep{Marino2009} NGC 1851 \citep{Han_2009}, Terzan 5 \citep{Ferraro_2009}, NGC 2419 \citep{Cohen2010}, NGC 3201 and \citep{Simmerer2013} in the Milky Way and G1 \citep{Meylan_2001} and
G78, G213, G280 \citep{Fuentes_Carrera_2008} in M31. Our results predict that some stripped nuclei which are lower than the typically defined minimum mass of M~$\approx$~2~$\times$~10\textsuperscript{6}~\(\textup{M}_\odot\) for UCDs \citep[e.g.][]{hilker2015} should exist, and that massive galaxies should contain a minimum of a few hundred, and massive clusters a few thousand, stripped nuclei around the range of masses that are typical of globular clusters. This is small in comparison to the several thousand globular clusters these massive galaxies are seen to host, and the $67300 \pm 1440$ globular clusters Virgo hosts \citep{Durrell2014}. However, since even low mass galaxy nuclei can contain black holes, the presence of a black hole with an unusually high mass might be used to discriminate between stripped nuclei and GCs, and investigate the population of stripped nuclei mingling with globular clusters. In a future work, we plan to predict the ages, metallicities and colours of stripped nuclei, so we will be able to predict the properties these stripped nuclei globular clusters will have, which will aid in finding them in observations.

\subsection{Does the UCD population depend more on central galaxy mass or host cluster mass?}

We find that the numbers of UCDs around individual galaxies are more strongly correlated with the stellar mass or halo mass of the central galaxy than they are with the mass of the host cluster. Fig.~\ref{fig:nosnlinearfits}, and Fig.~\ref{fig:multimassiveprojectedwithvirgo} respectively plot the merger tree and aperture \hl{sampled} average numbers of stripped nuclei per galaxy against stellar mass, while Fig.~\ref{fig:multimassiveprojectedwithvirgov2} plots aperture \hl{sampled} numbers against halo mass. The merger tree distribution shows a clear linear trend between the two, while the aperture \hl{sampled} plots show more scatter, but still a linear trend. For all graphs, there is no particular relationship between the host cluster and the number of stripped nuclei surrounding galaxies within the cluster, with the y-intercepts of the individual fits to each cluster plotted with the same slope falling well within one sigma of the y-intercept of the fit to all the clusters.  

Likely reasons for the correlation between stripped nuclei numbers and halo mass are that the number of mergers a galaxy undergoes will scale with the halo mass of the galaxy because of hierarchical formation. The correlation between stellar mass and numbers of stripped nuclei then follows on from the stellar mass-halo mass relation, which becomes steep at high halo masses. This results in an increase in scatter for the stellar mass-stripped nuclei number relation, as galaxies of a given stellar mass may inhabit a large range of halo masses. 

While Fig.~\ref{fig:clustnummasseslobf} and Fig.~\ref{fig:clustnummasseslobfv2} show a trend between cluster mass and stripped nuclei numbers this likely follows on from the galaxy halo mass-stripped nuclei number relation, rather than the mass of the cluster influencing the numbers of stripped nuclei the galaxies within it form. 

This result is, however, limited to  M~>~1~$\times$~10\textsuperscript{14}~\(\textup{M}_\odot\) clusters, and clusters with masses lower than this may have a different relationship between cluster mass and the number of stripped nuclei. \citet{Pfeffer2014} investigated clusters in the mass range of 1~$\times$~10\textsuperscript{13}~\(\textup{M}_\odot\) < M < 1~$\times$~10\textsuperscript{15}~\(\textup{M}_\odot\) and found that lower mass clusters are slightly more efficient in producing stripped nuclei than higher mass ones, and our narrow range of cluster masses makes it difficult to say for sure that there is no dependence on cluster mass for the number of stripped nuclei.
As UCDs have been detected around isolated galaxies that are not found within clusters, it would be worthwhile to plot them on the galaxy mass-UCD number relation to test if they are consistent with the number of UCDs found in galaxies in galaxy clusters. Another test could be holding the central galaxy mass constant while comparing different halo masses, however, this would be difficult due to the correlation between central galaxy mass and halo masses.

Fig.~\ref{fig:cumdistcnetralclusters} depicts the radial distribution of stripped nuclei in central galaxies in clusters of different masses and finds no particular correlation between radial distribution and cluster mass. There is no particular pattern between cluster mass and the shallowness or steepness of the distributions: for example, the galaxy within the most massive cluster is in the centre of the distributions. 

\subsection{Do objects transitioning from galaxies to UCDs by the stripping process exist?}
\citet{Pfeffer2014} predicted that Virgo should contain 9-12 disrupting nucleated galaxies. A number of studies have found galaxies that appear to be in the process of being stripped to become UCD-like objects \citep{Richtler2005, brodie2011, LiuPeng2015}. We find that around $6.3 \pm 3.2$ such objects should exist in Virgo-sized clusters and between 1-3 for massive galaxies, consistent with \citep{Pfeffer2014}. Therefore our simulations show that objects can exist that are nuclei in the process of being stripped. An additional number of merging galaxies may exist that have suffered significant tidal stripping but are still considered by EAGLE to be separate galaxies, and resemble UCDs in the process of formation.

\subsection{Caveats}
One source of uncertainty lies in the scatter from the nucleus-to-galaxy mass ratio. We have assumed that the nuclei survive the galaxy merger, based on predictions from previous studies \citep[e.g.][]{Bekki2001, Bekki2003}. These studies model the formation of stripped nuclei from typical dwarf galaxies in the 10\textsuperscript{8}~\(\textup{M}_\odot\) to 10\textsuperscript{10}~\(\textup{M}_\odot\) mass range with typical nuclei sizes of 2~-~20 per cent the stellar envelope mass. 
However, due to the scatter in the nucleus-to-galaxy mass ratio, some of our lower mass stripped nuclei will have been produced by M < 10\textsuperscript{8}~\(\textup{M}_\odot\) galaxies with very high nuclear star cluster to galaxy stellar mass ratios. The effect the threshing process has on these types of galaxies has been less studied than the typical mass range and nuclear star cluster/galaxy mass ratio. If low mass galaxies with very high nuclear star cluster masses are more likely to be totally disrupted than the typical models, it would substantially decrease our predicted numbers of low mass stripped nuclei. \citet{Janssens2016} studied ultra-diffuse galaxies and suggested that they are easier to disrupt than typical galaxies, for example. 

Additionally, our method relies on the assumption that nuclei at high redshifts have the same nucleus mass and nucleation fraction relationships, as those observed today. If these relationships are different at high redshifts, it will significantly impact the numbers and mass relationships of stripped nuclei that form. 

Another challenge is that the observations we are using rely on sampling UCDs with radii > 11~pc, while our stripped nuclei samples are mass-based, and we are unable to determine the radii of our stripped nuclei. As a result, we may overestimate the number of M~>~1~$\times$~10\textsuperscript{7}~\(\textup{M}_\odot\) stripped nuclei relative to observations of UCDs, because we cannot be sure that none of our sample of stripped nuclei have radii < 11~pc.

Other sources of uncertainty may be induced by the simulation, and are outlined in Section \ref{simcorr}. The unstable behaviours galaxies exhibit during mergers can impact the masses of the resulting stripped nuclei. Because stripping can begin several snapshots before the merger event, the stellar mass of the merging galaxy is taken to be the maximum of that in the snapshots before merger. However, in mergers that exhibit mass switching between the progenitor and the central galaxy the true stellar mass of the progenitor is difficult to determine, and so is taken from the snapshot in which the central galaxy mass is at a maximum. This may result in the mass of the stripped nucleus being underestimated if this snapshot did not contain an accurate stellar mass before stripping. This means a small number (< 5 per cent) of our stripped nuclei may have underestimated masses.

\hl{At the lower limit of our galaxy stellar masses of \mbox{M~>~1~$\times$~10\textsuperscript{7}~\(\textup{M}_\odot\)} galaxies may be resolved by less than 10 stellar particles. This could affect two properties: (1) the number of low mass galaxies in the simulation, and (2) their orbits. The number of low mass galaxies in EAGLE is fairly accurate, as it is a calibrated property \mbox{\citep{Schaye2015}}. Galaxy orbits should also be fairly accurate as this property will be determined largely by the dark matter halo, which is resolved by as many as \mbox{$\approx$}1000 particles at a stellar mass of \mbox{M~>~1~$\times$~10\textsuperscript{7}~\(\textup{M}_\odot\)}. We could raise the lower mass limit to improve the reliability of low-mass galaxies in the sample, but this would induce a separate source of error as we do expect lower mass galaxies to produce stripped nuclei.}

\hl{There is evidence that accretion of ex-situ material onto galaxies in the EAGLE simulation might be less efficient in high-mass clusters \mbox{\citep{Davison2020}}. This is possibly because high passing velocities inhibit true merging at z=0. This could affect the number of stripped nuclei formed and their masses. Extending our analysis to lower mass clusters and groups could help to verify our results. This we defer to a later paper.} 

It should be noted that there are large uncertainties in our numbers of stripped nuclei per galaxy (as seen in Fig.~\ref{fig:multimassiveprojectedwithvirgo}), and unknown uncertainties in the masses and numbers of observed UCDs caused by factors such as the radius limit and the use of a broad mass-to-light ratio. Due to this our results comparing the numbers and masses of stripped nuclei to UCDs are significantly less reliable than our radial distribution comparisons, which we are confident in.


\section{Summary}
In this paper, we present the first work to simulate stripped nuclei formation in hydrodynamical cosmological simulations and compare them to observations of UCDs. We use the EAGLE simulations to predict the numbers, masses and distributions of stripped nuclei formed in Virgo-sized galaxy clusters. Our conclusions are summarised as follows. 
\subsection{Radial distributions }
\begin{enumerate}
  \item Simulated stripped nuclei cluster strongly around the galaxies their progenitors merged with, rather than being spread evenly throughout the cluster. Observed UCDs similarly cluster around the most massive galaxies.
  \item The most viable method of comparing radial distributions of stripped nuclei involves using M~>~1~$\times$~10\textsuperscript{7}~\(\textup{M}_\odot\) stripped nuclei and rescaling the distributions by the galaxy effective radius. 
  \item The radial distribution of UCDs around M87 is consistent with the radial distribution of stripped nuclei around the central galaxy of the most massive simulated galaxy cluster for M~>~2~$\times$~10\textsuperscript{6}~\(\textup{M}_\odot\) stripped nuclei. 
  \item The radial distribution of UCDs around massive galaxies in the Virgo cluster are consistent with the radial distribution of simulated galaxies. 
\end{enumerate}

\subsection{Numbers and masses}
\begin{enumerate}
  \item Stripped nuclei can fully explain the number of M~>~1~$\times$~10\textsuperscript{7}~\(\textup{M}_\odot\) UCDs in the Virgo cluster around individual galaxies and in the full cluster.
  \item Galaxy halo mass predicts the number of UCDs surrounding the Virgo galaxies more accurately than stellar mass.
  \item We predict that there should exist a number of M~<~2~$\times$~10\textsuperscript{6}~\(\textup{M}_\odot\) stripped nuclei; several hundred surrounding massive galaxies and over a thousand in massive clusters.
\end{enumerate}

\subsection{Central galaxy mass vs host cluster mass}
\begin{enumerate}
    \item The numbers of UCDs around individual galaxies are strongly correlated with the stellar mass and halo mass of the central galaxy. This relationship does not depend on the mass of the host cluster.
    \item The radial distribution of stripped nuclei around central galaxies in different simulated clusters shows no dependence on cluster mass.
\end{enumerate}

\subsection{Transition objects}
\begin{enumerate}
  \item In Virgo-sized clusters there should exist around 6 objects transitioning from galaxies to stripped nuclei. Around 1-3 of these objects should exist for massive galaxies. 
\end{enumerate}

In conclusion, stripped nuclei can explain the numbers and distributions of high-mass UCDs in the Virgo cluster and around massive galaxies. The use of hydrodynamical simulations appears to solve some of the issues dark matter only simulations face with predicting numbers and distributions. We additionally predict that there should exist stripped nuclei with masses typical of globular clusters, as well as a number of transition objects.

\section*{Acknowledgements}
\hl{We thank the referee for the many helpful suggestions that improved this article.}

This work is based on observations obtained with MegaPrime/MegaCam, a joint project of CFHT and CEA/DAPNIA, at the Canada France Hawaii Telescope (CFHT) which is operated by the National Research Council (NRC) of Canada, the Institut National des Sciences de Univers of the Centre National de la Recherche Scientique (CNRS) of France, and the University of Hawaii. This work is based in part on data products produced at Terapix available at the Canadian Astronomy Data Centre as part of the CanadaFranceHawaii Telescope Legacy Survey, a collaborative project of NRC and CNRS.

We thank Lilian Garratt-Smithson for her assistance with the EAGLE database.

JP gratefully acknowledges funding from a European Research Council consolidator grant
(ERC-CoG-646928-Multi-Pop).

C.L. acknowledges support from the National Natural Science Foundation of China (NSFC, Grant No. 11673017, 11833005, 11933003, 11973033 and 11203017)

This work used the DiRAC@Durham facility managed by the Institute for Computational Cosmology on behalf of the STFC DiRAC HPC Facility (www.dirac.ac.uk). The equipment was funded by BEIS capital funding via STFC capital grants ST/K00042X/1, ST/P002293/1, ST/R002371/1 and ST/S002502/1, Durham University and STFC operations grant ST/R000832/1. DiRAC is part of the National e-Infrastructure.

\section*{DATA AVAILABILITY STATEMENT}

Data available on request.





\bibliographystyle{mnras}
\bibliography{biblio} 

\begin{thebibliography}{}
\makeatletter
\relax
\def\mn@urlcharsother{\let\do\@makeother \do\$\do\&\do\#\do\^\do\_\do\%\do\~}
\def\mn@doi{\begingroup\mn@urlcharsother \@ifnextchar [ {\mn@doi@}
  {\mn@doi@[]}}
\def\mn@doi@[#1]#2{\def\@tempa{#1}\ifx\@tempa\@empty \href
  {http://dx.doi.org/#2} {doi:#2}\else \href {http://dx.doi.org/#2} {#1}\fi
  \endgroup}
\def\mn@eprint#1#2{\mn@eprint@#1:#2::\@nil}
\def\mn@eprint@arXiv#1{\href {http://arxiv.org/abs/#1} {{\tt arXiv:#1}}}
\def\mn@eprint@dblp#1{\href {http://dblp.uni-trier.de/rec/bibtex/#1.xml}
  {dblp:#1}}
\def\mn@eprint@#1:#2:#3:#4\@nil{\def\@tempa {#1}\def\@tempb {#2}\def\@tempc
  {#3}\ifx \@tempc \@empty \let \@tempc \@tempb \let \@tempb \@tempa \fi \ifx
  \@tempb \@empty \def\@tempb {arXiv}\fi \@ifundefined
  {mn@eprint@\@tempb}{\@tempb:\@tempc}{\expandafter \expandafter \csname
  mn@eprint@\@tempb\endcsname \expandafter{\@tempc}}}

\bibitem[\protect\citeauthoryear{Ahn et~al.,}{Ahn et~al.}{2017}]{Ahn2017}
Ahn C.~P.,  et~al., 2017, \mn@doi [\apj] {10.3847/1538-4357/aa6972}, 839, 72

\bibitem[\protect\citeauthoryear{Ahn et~al.}{Ahn et~al.}{2018}]{Ahn2018}
Ahn C.~P.,  et~al., 2018, \mn@doi [\apj] {10.3847/1538-4357/aabc57}, \href
  {http://adsabs.harvard.edu/abs/2018ApJ...858..102A} {858, 102}

\bibitem[\protect\citeauthoryear{{Bassino}, {Muzzio}  \& {Rabolli}}{{Bassino}
  et~al.}{1994}]{Bassino1994}
{Bassino} L.~P.,  {Muzzio} J.~C.,   {Rabolli} M.,  1994, \mn@doi [\apj]
  {10.1086/174514}, \href {http://adsabs.harvard.edu/abs/1994ApJ...431..634B}
  {431, 634}

\bibitem[\protect\citeauthoryear{{Bekki}, {Couch}  \& {Drinkwater}}{{Bekki}
  et~al.}{2001}]{Bekki2001}
{Bekki} K.,  {Couch} W.~J.,   {Drinkwater} M.~J.,  2001, \mn@doi [\apjl]
  {10.1086/320339}, \href {http://adsabs.harvard.edu/abs/2001ApJ...552L.105B}
  {552, L105}

\bibitem[\protect\citeauthoryear{Bekki et~al.}{Bekki et~al.}{2003}]{Bekki2003}
Bekki K.,  et~al., 2003, \mn@doi [\mnras] {10.1046/j.1365-8711.2003.06916.x},
  \href {http://adsabs.harvard.edu/abs/2003MNRAS.344..399B} {344, 399}

\bibitem[\protect\citeauthoryear{{Binney} \& {Tremaine}}{{Binney} \&
  {Tremaine}}{1987}]{Binney1987}
{Binney} J.,  {Tremaine} S.,  1987, {Galactic dynamics}.
Princeton University Press

\bibitem[\protect\citeauthoryear{{Boylan-Kolchin}, {Springel}, {White},
  {Jenkins}  \& {Lemson}}{{Boylan-Kolchin} et~al.}{2009}]{Boylan2009}
{Boylan-Kolchin} M.,  {Springel} V.,  {White} S. D.~M.,  {Jenkins} A.,
  {Lemson} G.,  2009, \mn@doi [\mnras] {10.1111/j.1365-2966.2009.15191.x},
  \href {https://ui.adsabs.harvard.edu/abs/2009MNRAS.398.1150B} {398, 1150}

\bibitem[\protect\citeauthoryear{{Brodie}, {Romanowsky}, {Strader}  \&
  {Forbes}}{{Brodie} et~al.}{2011}]{brodie2011}
{Brodie} J.~P.,  {Romanowsky} A.~J.,  {Strader} J.,   {Forbes} D.~A.,  2011,
  \mn@doi [\aj] {10.1088/0004-6256/142/6/199}, \href
  {https://ui.adsabs.harvard.edu/abs/2011AJ....142..199B} {142, 199}

\bibitem[\protect\citeauthoryear{{Br{\"u}ns} \& {Kroupa}}{{Br{\"u}ns} \&
  {Kroupa}}{2012}]{brun2012}
{Br{\"u}ns} R.~C.,  {Kroupa} P.,  2012, \mn@doi [\aap]
  {10.1051/0004-6361/201219693}, 547, A65

\bibitem[\protect\citeauthoryear{{Br{\"u}ns}, {Kroupa}, {Fellhauer}, {Metz}  \&
  {Assmann}}{{Br{\"u}ns} et~al.}{2011}]{brun2011}
{Br{\"u}ns} R.~C.,  {Kroupa} P.,  {Fellhauer} M.,  {Metz} M.,   {Assmann} P.,
  2011, \mn@doi [\aap] {10.1051/0004-6361/201016220}, 529, A138

\bibitem[\protect\citeauthoryear{{Carretta} et~al.,}{{Carretta}
  et~al.}{2010}]{Carretta2010}
{Carretta} E.,  et~al., 2010, \mn@doi [\apjl] {10.1088/2041-8205/714/1/L7},
  \href {https://ui.adsabs.harvard.edu/abs/2010ApJ...714L...7C} {714, L7}

\bibitem[\protect\citeauthoryear{{Caso}, {Bassino}, {Richtler}, {Smith
  Castelli}  \& {Faifer}}{{Caso} et~al.}{2013}]{caso2013}
{Caso} J.~P.,  {Bassino} L.~P.,  {Richtler} T.,  {Smith Castelli} A.~V.,
  {Faifer} F.~R.,  2013, \mn@doi [\mnras] {10.1093/mnras/sts687}, \href
  {https://ui.adsabs.harvard.edu/abs/2013MNRAS.430.1088C} {430, 1088}

\bibitem[\protect\citeauthoryear{{Casteels} et~al.,}{{Casteels}
  et~al.}{2014}]{Casteels2014}
{Casteels} K. R.~V.,  et~al., 2014, \mn@doi [\mnras] {10.1093/mnras/stu1799},
  \href {https://ui.adsabs.harvard.edu/abs/2014MNRAS.445.1157C} {445, 1157}

\bibitem[\protect\citeauthoryear{{Chiboucas} et~al.,}{{Chiboucas}
  et~al.}{2011}]{Chiboucas2011}
{Chiboucas} K.,  et~al., 2011, \mn@doi [\apj] {10.1088/0004-637X/737/2/86},
  \href {https://ui.adsabs.harvard.edu/abs/2011ApJ...737...86C} {737, 86}

\bibitem[\protect\citeauthoryear{{Chilingarian} \& {Mamon}}{{Chilingarian} \&
  {Mamon}}{2008}]{Chilingarian2008}
{Chilingarian} I.~V.,  {Mamon} G.~A.,  2008, \mn@doi [\mnras]
  {10.1111/j.1745-3933.2008.00438.x}, \href
  {https://ui.adsabs.harvard.edu/abs/2008MNRAS.385L..83C} {385, L83}

\bibitem[\protect\citeauthoryear{{Chilingarian}, {Mieske}, {Hilker}  \&
  {Infante}}{{Chilingarian} et~al.}{2011}]{Chilingarian2011}
{Chilingarian} I.~V.,  {Mieske} S.,  {Hilker} M.,   {Infante} L.,  2011,
  \mn@doi [\mnras] {10.1111/j.1365-2966.2010.18000.x}, \href
  {https://ui.adsabs.harvard.edu/abs/2011MNRAS.412.1627C} {412, 1627}

\bibitem[\protect\citeauthoryear{Clauwens, Schaye, Franx  \& Bower}{Clauwens
  et~al.}{2018}]{Clauwens_2018}
Clauwens B.,  Schaye J.,  Franx M.,   Bower R.~G.,  2018, \mn@doi [\mnras]
  {10.1093/mnras/sty1229}, 478, 3994

\bibitem[\protect\citeauthoryear{Cohen, Kirby, Simon  \& Geha}{Cohen
  et~al.}{2010}]{Cohen2010}
Cohen J.,  Kirby E.,  Simon J.,   Geha M.,  2010, \mn@doi [\apj]
  {10.1088/0004-637X/725/1/288}, 725

\bibitem[\protect\citeauthoryear{Cote, McLaughlin, Cohen  \& Blakeslee}{Cote
  et~al.}{2003}]{Cote_2003}
Cote P.,  McLaughlin D.~E.,  Cohen J.~G.,   Blakeslee J.~P.,  2003, \mn@doi
  [\apj] {10.1086/375488}, 591, 850

\bibitem[\protect\citeauthoryear{C{\^o}t{\'e} et~al.}{C{\^o}t{\'e}
  et~al.}{2006}]{Cote2006}
C{\^o}t{\'e} P.,  et~al., 2006, \mn@doi [\apjs] {10.1086/504042}, \href
  {http://adsabs.harvard.edu/abs/2006ApJS..165...57C} {165, 57}

\bibitem[\protect\citeauthoryear{Crain et~al.,}{Crain
  et~al.}{2015}]{Crain_2015}
Crain R.~A.,  et~al., 2015, \mn@doi [\mnras] {10.1093/mnras/stv725}, 450, 1937

\bibitem[\protect\citeauthoryear{{Da Rocha}, {Mieske}, {Georgiev}, {Hilker},
  {Ziegler}  \& {Mendes de Oliveira}}{{Da Rocha} et~al.}{2011}]{DaRocha2011}
{Da Rocha} C.,  {Mieske} S.,  {Georgiev} I.~Y.,  {Hilker} M.,  {Ziegler} B.~L.,
    {Mendes de Oliveira} C.,  2011, \mn@doi [\aap]
  {10.1051/0004-6361/201015353}, 525, A86

\bibitem[\protect\citeauthoryear{{Davis}, {Efstathiou}, {Frenk}  \&
  {White}}{{Davis} et~al.}{1985}]{Davis1985}
{Davis} M.,  {Efstathiou} G.,  {Frenk} C.~S.,   {White} S.~D.~M.,  1985,
  \mn@doi [\apj] {10.1086/163168}, \href
  {https://ui.adsabs.harvard.edu/abs/1985ApJ...292..371D} {292, 371}

\bibitem[\protect\citeauthoryear{{Davison}, {Norris}, {Pfeffer}, {Davies}  \&
  {Crain}}{{Davison} et~al.}{2020}]{Davison2020}
{Davison} T.~A.,  {Norris} M.~A.,  {Pfeffer} J.~L.,  {Davies} J.~J.,   {Crain}
  R.~A.,  2020, \mn@doi [\mnras] {10.1093/mnras/staa1816}, \href
  {https://ui.adsabs.harvard.edu/abs/2020MNRAS.497...81D} {497, 81}

\bibitem[\protect\citeauthoryear{{Dolag}, {Borgani}, {Murante}  \&
  {Springel}}{{Dolag} et~al.}{2009}]{Dolag2009}
{Dolag} K.,  {Borgani} S.,  {Murante} G.,   {Springel} V.,  2009, \mn@doi
  [\mnras] {10.1111/j.1365-2966.2009.15034.x}, \href
  {https://ui.adsabs.harvard.edu/abs/2009MNRAS.399..497D} {399, 497}

\bibitem[\protect\citeauthoryear{{Drinkwater}, {Jones}, {Gregg}  \&
  {Phillipps}}{{Drinkwater} et~al.}{2000}]{Drinkwater2000}
{Drinkwater} M.~J.,  {Jones} J.~B.,  {Gregg} M.~D.,   {Phillipps} S.,  2000,
  \mn@doi [\pasa] {10.1071/AS00034}, \href
  {https://ui.adsabs.harvard.edu/abs/2000PASA...17..227D} {17, 227}

\bibitem[\protect\citeauthoryear{Drinkwater et~al.}{Drinkwater
  et~al.}{2003}]{Drinkwater2003}
Drinkwater M.~J.,  et~al., 2003, \mn@doi [\nat] {10.1038/nature01666}, \href
  {http://adsabs.harvard.edu/abs/2003Natur.423..519D} {423, 519}

\bibitem[\protect\citeauthoryear{{Drinkwater}, {Gregg}, {Couch}, {Ferguson},
  {Hilker}, {Jones}, {Karick}  \& {Phillipps}}{{Drinkwater}
  et~al.}{2004}]{Drinkwater2004}
{Drinkwater} M.~J.,  {Gregg} M.~D.,  {Couch} W.~J.,  {Ferguson} H.~C.,
  {Hilker} M.,  {Jones} J.~B.,  {Karick} A.,   {Phillipps} S.,  2004, \mn@doi
  [\pasa] {10.1071/AS04048}, \href
  {https://ui.adsabs.harvard.edu/abs/2004PASA...21..375D} {21, 375}

\bibitem[\protect\citeauthoryear{{Durrell} et~al.,}{{Durrell}
  et~al.}{2014}]{Durrell2014}
{Durrell} P.~R.,  et~al., 2014, \mn@doi [\apj] {10.1088/0004-637X/794/2/103},
  \href {https://ui.adsabs.harvard.edu/abs/2014ApJ...794..103D} {794, 103}

\bibitem[\protect\citeauthoryear{{Evstigneeva}, {Drinkwater}, {Jurek}, {Firth},
  {Jones}, {Gregg}  \& {Phillipps}}{{Evstigneeva} et~al.}{2007}]{Evstig2007a}
{Evstigneeva} E.~A.,  {Drinkwater} M.~J.,  {Jurek} R.,  {Firth} P.,  {Jones}
  J.~B.,  {Gregg} M.~D.,   {Phillipps} S.,  2007, \mn@doi [\mnras]
  {10.1111/j.1365-2966.2007.11856.x}, \href
  {https://ui.adsabs.harvard.edu/abs/2007MNRAS.378.1036E} {378, 1036}

\bibitem[\protect\citeauthoryear{Evstigneeva et~al.,}{Evstigneeva
  et~al.}{2008}]{Evstig2008A}
Evstigneeva E.~A.,  et~al., 2008, \mn@doi [\aj] {10.1088/0004-6256/136/1/461},
  136, 461–478

\bibitem[\protect\citeauthoryear{{Fellhauer} \& {Kroupa}}{{Fellhauer} \&
  {Kroupa}}{2002}]{Fellhauer2002}
{Fellhauer} M.,  {Kroupa} P.,  2002, \mn@doi [\mnras]
  {10.1046/j.1365-8711.2002.05087.x}, \href
  {https://ui.adsabs.harvard.edu/abs/2002MNRAS.330..642F} {330, 642}

\bibitem[\protect\citeauthoryear{{Ferrarese} et~al.,}{{Ferrarese}
  et~al.}{2012}]{Ferrarese2012}
{Ferrarese} L.,  et~al., 2012, \mn@doi [\apjs] {10.1088/0067-0049/200/1/4},
  \href {https://ui.adsabs.harvard.edu/abs/2012ApJS..200....4F} {200, 4}

\bibitem[\protect\citeauthoryear{Ferraro et~al.,}{Ferraro
  et~al.}{2009}]{Ferraro_2009}
Ferraro F.~R.,  et~al., 2009, \mn@doi [Nature] {10.1038/nature08581}, 462, 483

\bibitem[\protect\citeauthoryear{{Forbes} et~al.,}{{Forbes}
  et~al.}{2004}]{Forbes2004}
{Forbes} D.~A.,  et~al., 2004, \mn@doi [\mnras]
  {10.1111/j.1365-2966.2004.08333.x}, \href
  {https://ui.adsabs.harvard.edu/abs/2004MNRAS.355..608F} {355, 608}

\bibitem[\protect\citeauthoryear{Forbes, Pota, Usher, Strader, Romanowsky,
  Brodie, Arnold  \& Spitler}{Forbes et~al.}{2013}]{Forbes2013}
Forbes D.,  Pota V.,  Usher C.,  Strader J.,  Romanowsky A.,  Brodie J.,
  Arnold J.,   Spitler L.,  2013, \mn@doi [\mnras] {10.1093/mnrasl/slt078},
  435, L6

\bibitem[\protect\citeauthoryear{{Francis}, {Drinkwater}, {Chilingarian},
  {Bolt}  \& {Firth}}{{Francis} et~al.}{2012}]{Francis2012}
{Francis} K.~J.,  {Drinkwater} M.~J.,  {Chilingarian} I.~V.,  {Bolt} A.~M.,
  {Firth} P.,  2012, \mn@doi [\mnras] {10.1111/j.1365-2966.2012.21465.x}, \href
  {https://ui.adsabs.harvard.edu/abs/2012MNRAS.425..325F} {425, 325}

\bibitem[\protect\citeauthoryear{Fuentes-Carrera, Jablonka, Sarajedini,
  Bridges, Djorgovski  \& Meylan}{Fuentes-Carrera
  et~al.}{2008}]{Fuentes_Carrera_2008}
Fuentes-Carrera I.,  Jablonka P.,  Sarajedini A.,  Bridges T.,  Djorgovski G.,
   Meylan G.,  2008, \mn@doi [\aap] {10.1051/0004-6361:20078899}, 483, 769

\bibitem[\protect\citeauthoryear{{Furlong} et~al.,}{{Furlong}
  et~al.}{2015}]{Furlong2015}
{Furlong} M.,  et~al., 2015, \mn@doi [\mnras] {10.1093/mnras/stv852}, \href
  {https://ui.adsabs.harvard.edu/abs/2015MNRAS.450.4486F} {450, 4486}

\bibitem[\protect\citeauthoryear{{Furlong} et~al.,}{{Furlong}
  et~al.}{2017}]{Furlong2017}
{Furlong} M.,  et~al., 2017, \mn@doi [\mnras] {10.1093/mnras/stw2740}, \href
  {https://ui.adsabs.harvard.edu/abs/2017MNRAS.465..722F} {465, 722}

\bibitem[\protect\citeauthoryear{{Galianni}, {Patat}, {Higdon}, {Mieske}  \&
  {Kroupa}}{{Galianni} et~al.}{2010}]{Galianni2010}
{Galianni} P.,  {Patat} F.,  {Higdon} J.~L.,  {Mieske} S.,   {Kroupa} P.,
  2010, \mn@doi [\aap] {10.1051/0004-6361/200913518}, \href
  {https://ui.adsabs.harvard.edu/abs/2010A&A...521A..20G} {521, A20}

\bibitem[\protect\citeauthoryear{Gebhardt \& Thomas}{Gebhardt \&
  Thomas}{2009}]{Gebhardt_2009}
Gebhardt K.,  Thomas J.,  2009, \mn@doi [\apj] {10.1088/0004-637x/700/2/1690},
  700, 1690–1701

\bibitem[\protect\citeauthoryear{{Goerdt}, {Moore}, {Kazantzidis}, {Kaufmann},
  {Macci{\`o}}  \& {Stadel}}{{Goerdt} et~al.}{2008}]{Goerdt2008}
{Goerdt} T.,  {Moore} B.,  {Kazantzidis} S.,  {Kaufmann} T.,  {Macci{\`o}}
  A.~V.,   {Stadel} J.,  2008, \mn@doi [\mnras]
  {10.1111/j.1365-2966.2008.12982.x}, \href
  {https://ui.adsabs.harvard.edu/abs/2008MNRAS.385.2136G} {385, 2136}

\bibitem[\protect\citeauthoryear{{Guo} et~al.,}{{Guo} et~al.}{2011}]{Guo2011}
{Guo} Q.,  et~al., 2011, \mn@doi [\mnras] {10.1111/j.1365-2966.2010.18114.x},
  \href {https://ui.adsabs.harvard.edu/abs/2011MNRAS.413..101G} {413, 101}

\bibitem[\protect\citeauthoryear{{Ha{\c s}egan} et~al.}{{Ha{\c s}egan}
  et~al.}{2005}]{has2005}
{Ha{\c s}egan} M.,  et~al., 2005, \mn@doi [\apj] {10.1086/430342}, \href
  {http://adsabs.harvard.edu/abs/2005ApJ...627..203H} {627, 203}

\bibitem[\protect\citeauthoryear{Han, Lee, Joo, Sohn, Yoon, Kim  \& Lee}{Han
  et~al.}{2009}]{Han_2009}
Han S.-I.,  Lee Y.-W.,  Joo S.-J.,  Sohn S.~T.,  Yoon S.-J.,  Kim H.-S.,   Lee
  J.-W.,  2009, \mn@doi [\apj] {10.1088/0004-637x/707/2/l190}, 707, L190

\bibitem[\protect\citeauthoryear{{Hau}, {Spitler}, {Forbes}, {Proctor},
  {Strader}, {Mendel}, {Brodie}  \& {Harris}}{{Hau} et~al.}{2009}]{Hau200}
{Hau} G. K.~T.,  {Spitler} L.~R.,  {Forbes} D.~A.,  {Proctor} R.~N.,  {Strader}
  J.,  {Mendel} J.~T.,  {Brodie} J.~P.,   {Harris} W.~E.,  2009, \mn@doi
  [\mnras] {10.1111/j.1745-3933.2009.00618.x}, \href
  {https://ui.adsabs.harvard.edu/abs/2009MNRAS.394L..97H} {394, L97}

\bibitem[\protect\citeauthoryear{Hilker}{Hilker}{2015}]{hilker2015}
Hilker M.,  2015, Stellar population properties of the most massive globular
  clusters and ultra-compact dwarf galaxies of the Fornax cluster (\mn@eprint
  {arXiv} {1512.07414})

\bibitem[\protect\citeauthoryear{{Hilker} \& {Richtler}}{{Hilker} \&
  {Richtler}}{2000}]{hilker2000}
{Hilker} M.,  {Richtler} T.,  2000, \aap, \href
  {https://ui.adsabs.harvard.edu/abs/2000A&A...362..895H} {362, 895}

\bibitem[\protect\citeauthoryear{{Hilker}, {Infante}, {Vieira}, {Kissler-Patig}
   \& {Richtler}}{{Hilker} et~al.}{1999}]{Hilker1999}
{Hilker} M.,  {Infante} L.,  {Vieira} G.,  {Kissler-Patig} M.,   {Richtler} T.,
   1999, \mn@doi [\aaps] {10.1051/aas:1999434}, 134, 75

\bibitem[\protect\citeauthoryear{{Hilker}, {Mieske}, {Baumgardt}  \&
  {Dabringhausen}}{{Hilker} et~al.}{2008}]{Hilker2008}
{Hilker} M.,  {Mieske} S.,  {Baumgardt} H.,   {Dabringhausen} J.,  2008, in
  {Vesperini} E.,  {Giersz} M.,   {Sills} A.,  eds,  IAU Symposium Vol. 246,
  Dynamical Evolution of Dense Stellar Systems. pp 427--428,
  \mn@doi{10.1017/S1743921308016104}

\bibitem[\protect\citeauthoryear{Huxor, Phillipps, Price  \& Harniman}{Huxor
  et~al.}{2011}]{Huxor_2011}
Huxor A.~P.,  Phillipps S.,  Price J.,   Harniman R.,  2011, \mn@doi [\mnras]
  {10.1111/j.1365-2966.2011.18655.x}, 414, 3557

\bibitem[\protect\citeauthoryear{Hwang et~al.,}{Hwang
  et~al.}{2008}]{Hwang_2008}
Hwang H.~S.,  et~al., 2008, \mn@doi [\apj] {10.1086/524001}, 674, 869

\bibitem[\protect\citeauthoryear{{Ibata}, {Gilmore}  \& {Irwin}}{{Ibata}
  et~al.}{1994}]{Ibata1994}
{Ibata} R.~A.,  {Gilmore} G.,   {Irwin} M.~J.,  1994, \mn@doi [\nat]
  {10.1038/370194a0}, \href
  {https://ui.adsabs.harvard.edu/abs/1994Natur.370..194I} {370, 194}

\bibitem[\protect\citeauthoryear{Janssens, Abraham, Brodie, Forbes, Romanowsky
  \& Dokkum}{Janssens et~al.}{2016}]{Janssens2016}
Janssens S.,  Abraham R.,  Brodie J.,  Forbes D.,  Romanowsky A.,   Dokkum P.,
  2016, \mn@doi [\apjl] {10.3847/2041-8213/aa667d}, 839

\bibitem[\protect\citeauthoryear{Janz et~al.,}{Janz et~al.}{2015}]{Janz_2015}
Janz J.,  et~al., 2015, \mn@doi [\mnras] {10.1093/mnras/stv2636}, 456, 617

\bibitem[\protect\citeauthoryear{{Jennings} et~al.,}{{Jennings}
  et~al.}{2015}]{Jennings2015}
{Jennings} Z.~G.,  et~al., 2015, \mn@doi [\apjl] {10.1088/2041-8205/812/1/L10},
  \href {https://ui.adsabs.harvard.edu/abs/2015ApJ...812L..10J} {812, L10}

\bibitem[\protect\citeauthoryear{{Kroupa}}{{Kroupa}}{1998}]{Kroupa1998}
{Kroupa} P.,  1998, \mn@doi [\mnras] {10.1046/j.1365-8711.1998.01892.x}, \href
  {https://ui.adsabs.harvard.edu/abs/1998MNRAS.300..200K} {300, 200}

\bibitem[\protect\citeauthoryear{{Lacey} \& {Cole}}{{Lacey} \&
  {Cole}}{1993}]{Lacey1993}
{Lacey} C.,  {Cole} S.,  1993, \mn@doi [\mnras] {10.1093/mnras/262.3.627},
  \href {https://ui.adsabs.harvard.edu/abs/1993MNRAS.262..627L} {262, 627}

\bibitem[\protect\citeauthoryear{{Lagos} et~al.,}{{Lagos}
  et~al.}{2015}]{Lagos2015}
{Lagos} C. d.~P.,  et~al., 2015, \mn@doi [\mnras] {10.1093/mnras/stv1488},
  \href {https://ui.adsabs.harvard.edu/abs/2015MNRAS.452.3815L} {452, 3815}

\bibitem[\protect\citeauthoryear{Lee, Joo, Sohn, Rey, Lee  \& Walker}{Lee
  et~al.}{1999}]{Lee1999}
Lee Y.-W.,  Joo J.-M.,  Sohn Y.-J.,  Rey S.-C.,  Lee H.-c.,   Walker A.,  1999,
  \mn@doi [Nature] {10.1038/46985}, 402

\bibitem[\protect\citeauthoryear{{Liu} et~al.}{{Liu} et~al.}{2015a}]{Liu2015}
{Liu} C.,  et~al., 2015a, \mn@doi [\apj] {10.1088/0004-637X/812/1/34}, \href
  {http://adsabs.harvard.edu/abs/2015ApJ...812...34L} {812, 34}

\bibitem[\protect\citeauthoryear{{Liu} et~al.,}{{Liu}
  et~al.}{2015b}]{LiuPeng2015}
{Liu} C.,  et~al., 2015b, \mn@doi [\apjl] {10.1088/2041-8205/812/1/L2}, \href
  {https://ui.adsabs.harvard.edu/abs/2015ApJ...812L...2L} {812, L2}

\bibitem[\protect\citeauthoryear{Liu et~al.,}{Liu
  et~al.}{2020}]{liu2020generation}
Liu C.,  et~al., 2020, \apjs, 250

\bibitem[\protect\citeauthoryear{{Madrid} \& {Donzelli}}{{Madrid} \&
  {Donzelli}}{2013}]{Madrid2013}
{Madrid} J.~P.,  {Donzelli} C.~J.,  2013, \mn@doi [\apj]
  {10.1088/0004-637X/770/2/158}, \href
  {https://ui.adsabs.harvard.edu/abs/2013ApJ...770..158M} {770, 158}

\bibitem[\protect\citeauthoryear{{Madrid} et~al.,}{{Madrid}
  et~al.}{2010}]{Madrid2010}
{Madrid} J.~P.,  et~al., 2010, \mn@doi [\apj] {10.1088/0004-637X/722/2/1707},
  \href {https://ui.adsabs.harvard.edu/abs/2010ApJ...722.1707M} {722, 1707}

\bibitem[\protect\citeauthoryear{{Marino}, {Milone}, {Piotto}, {Villanova},
  {Bedin}, {Bellini}  \& {Renzini}}{{Marino} et~al.}{2009}]{Marino2009}
{Marino} A.~F.,  {Milone} A.~P.,  {Piotto} G.,  {Villanova} S.,  {Bedin} L.~R.,
   {Bellini} A.,   {Renzini} A.,  2009, \mn@doi [\aap]
  {10.1051/0004-6361/200911827}, \href
  {https://ui.adsabs.harvard.edu/abs/2009A&A...505.1099M} {505, 1099}

\bibitem[\protect\citeauthoryear{{McAlpine} et~al.,}{{McAlpine}
  et~al.}{2016}]{McAlpine2016}
{McAlpine} S.,  et~al., 2016, \mn@doi [Astronomy and Computing]
  {10.1016/j.ascom.2016.02.004}, \href
  {https://ui.adsabs.harvard.edu/abs/2016A&C....15...72M} {15, 72}

\bibitem[\protect\citeauthoryear{{McLaughlin}}{{McLaughlin}}{1999}]{McLaughlin1999}
{McLaughlin} D.~E.,  1999, \mn@doi [\apjl] {10.1086/311860}, \href
  {https://ui.adsabs.harvard.edu/abs/1999ApJ...512L...9M} {512, L9}

\bibitem[\protect\citeauthoryear{{Mei} et~al.,}{{Mei} et~al.}{2007}]{Mei2007}
{Mei} S.,  et~al., 2007, \mn@doi [\apj] {10.1086/509598}, \href
  {https://ui.adsabs.harvard.edu/abs/2007ApJ...655..144M} {655, 144}

\bibitem[\protect\citeauthoryear{Meylan, Sarajedini, Jablonka, Djorgovski,
  Bridges  \& Rich}{Meylan et~al.}{2001}]{Meylan_2001}
Meylan G.,  Sarajedini A.,  Jablonka P.,  Djorgovski S.~G.,  Bridges T.,   Rich
  R.~M.,  2001, \mn@doi [\aj] {10.1086/321166}, 122, 830

\bibitem[\protect\citeauthoryear{{Mieske} \& {Kroupa}}{{Mieske} \&
  {Kroupa}}{2008}]{Mieske2008}
{Mieske} S.,  {Kroupa} P.,  2008, \mn@doi [\apj] {10.1086/528739}, \href
  {http://adsabs.harvard.edu/abs/2008ApJ...677..276M} {677, 276}

\bibitem[\protect\citeauthoryear{{Mieske}, {Hilker}  \& {Infante}}{{Mieske}
  et~al.}{2002}]{Mieske2002}
{Mieske} S.,  {Hilker} M.,   {Infante} L.,  2002, \mn@doi [\aap]
  {10.1051/0004-6361:20011833}, 383, 823

\bibitem[\protect\citeauthoryear{{Mieske}, {Hilker}  \& {Infante}}{{Mieske}
  et~al.}{2004}]{Mieske2004}
{Mieske} S.,  {Hilker} M.,   {Infante} L.,  2004, \mn@doi [\aap]
  {10.1051/0004-6361:20035723}, 418, 445

\bibitem[\protect\citeauthoryear{{Mieske}, {Hilker}, {Infante}  \&
  {Jord{\'a}n}}{{Mieske} et~al.}{2006}]{Mieske2006}
{Mieske} S.,  {Hilker} M.,  {Infante} L.,   {Jord{\'a}n} A.,  2006, \mn@doi
  [\aj] {10.1086/500583}, \href
  {https://ui.adsabs.harvard.edu/abs/2006AJ....131.2442M} {131, 2442}

\bibitem[\protect\citeauthoryear{{Mieske}, {Hilker}, {Jord{\'a}n}, {Infante}
  \& {Kissler-Patig}}{{Mieske} et~al.}{2007}]{Mieske2007a}
{Mieske} S.,  {Hilker} M.,  {Jord{\'a}n} A.,  {Infante} L.,   {Kissler-Patig}
  M.,  2007, \mn@doi [\aap] {10.1051/0004-6361:20077631}, 472, 111

\bibitem[\protect\citeauthoryear{Mieske et~al.,}{Mieske
  et~al.}{2008}]{Mieske_2008}
Mieske S.,  et~al., 2008, \mn@doi [\aap] {10.1051/0004-6361:200810077}, 487,
  921

\bibitem[\protect\citeauthoryear{{Mieske}, {Hilker}  \& {Misgeld}}{{Mieske}
  et~al.}{2012}]{Mieske2012}
{Mieske} S.,  {Hilker} M.,   {Misgeld} I.,  2012, \mn@doi [\aap]
  {10.1051/0004-6361/201117634}, 537, A3

\bibitem[\protect\citeauthoryear{Mieske et~al.}{Mieske
  et~al.}{2013}]{Mieske2013}
Mieske S.,  et~al., 2013, \mn@doi [\aap] {10.1051/0004-6361/201322167}, 558,
  A14

\bibitem[\protect\citeauthoryear{{Misgeld}, {Mieske}, {Hilker}, {Richtler},
  {Georgiev}  \& {Schuberth}}{{Misgeld} et~al.}{2011}]{Misgeld2011}
{Misgeld} I.,  {Mieske} S.,  {Hilker} M.,  {Richtler} T.,  {Georgiev} I.~Y.,
  {Schuberth} Y.,  2011, \mn@doi [\aap] {10.1051/0004-6361/201116728}, 531, A4

\bibitem[\protect\citeauthoryear{{Norris} \& {Kannappan}}{{Norris} \&
  {Kannappan}}{2011}]{Norris2011}
{Norris} M.~A.,  {Kannappan} S.~J.,  2011, \mn@doi [\mnras]
  {10.1111/j.1365-2966.2011.18440.x}, \href
  {https://ui.adsabs.harvard.edu/abs/2011MNRAS.414..739N} {414, 739}

\bibitem[\protect\citeauthoryear{{Norris}, {Escudero}, {Faifer}, {Kannappan},
  {Forte}  \& {van den Bosch}}{{Norris} et~al.}{2015}]{Norris2015}
{Norris} M.~A.,  {Escudero} C.~G.,  {Faifer} F.~R.,  {Kannappan} S.~J.,
  {Forte} J.~C.,   {van den Bosch} R. C.~E.,  2015, \mn@doi [\mnras]
  {10.1093/mnras/stv1221}, \href
  {https://ui.adsabs.harvard.edu/abs/2015MNRAS.451.3615N} {451, 3615}

\bibitem[\protect\citeauthoryear{{Oldham} \& {Auger}}{{Oldham} \&
  {Auger}}{2016}]{Oldham2016}
{Oldham} L.~J.,  {Auger} M.~W.,  2016, \mn@doi [\mnras]
  {10.1093/mnras/stv2244}, \href
  {https://ui.adsabs.harvard.edu/abs/2016MNRAS.455..820O} {455, 820}

\bibitem[\protect\citeauthoryear{{Oser}, {Ostriker}, {Naab}, {Johansson}  \&
  {Burkert}}{{Oser} et~al.}{2010}]{Oser2010}
{Oser} L.,  {Ostriker} J.~P.,  {Naab} T.,  {Johansson} P.~H.,   {Burkert} A.,
  2010, \mn@doi [\apj] {10.1088/0004-637X/725/2/2312}, \href
  {https://ui.adsabs.harvard.edu/abs/2010ApJ...725.2312O} {725, 2312}

\bibitem[\protect\citeauthoryear{{Pfeffer} \& {Baumgardt}}{{Pfeffer} \&
  {Baumgardt}}{2013}]{Pfeffer2013}
{Pfeffer} J.,  {Baumgardt} H.,  2013, \mn@doi [\mnras] {10.1093/mnras/stt867},
  \href {http://adsabs.harvard.edu/abs/2013MNRAS.433.1997P} {433, 1997}

\bibitem[\protect\citeauthoryear{Pfeffer et~al.}{Pfeffer
  et~al.}{2014}]{Pfeffer2014}
Pfeffer J.,  et~al., 2014, \mn@doi [\mnras] {10.1093/mnras/stu1705}, \href
  {http://adsabs.harvard.edu/abs/2014MNRAS.444.3670P} {444, 3670}

\bibitem[\protect\citeauthoryear{Pfeffer, Hilker, Baumgardt  \&
  Griffen}{Pfeffer et~al.}{2016}]{Pfeffer2016}
Pfeffer J.,  Hilker M.,  Baumgardt H.,   Griffen B.~F.,  2016, \mn@doi [\mnras]
  {10.1093/mnras/stw498}, 458, 2492

\bibitem[\protect\citeauthoryear{{Qu} et~al.,}{{Qu} et~al.}{2017}]{Qu2017}
{Qu} Y.,  et~al., 2017, \mn@doi [\mnras] {10.1093/mnras/stw2437}, \href
  {https://ui.adsabs.harvard.edu/abs/2017MNRAS.464.1659Q} {464, 1659}

\bibitem[\protect\citeauthoryear{{Richtler}, {Dirsch}, {Larsen}, {Hilker}  \&
  {Infante}}{{Richtler} et~al.}{2005}]{Richtler2005}
{Richtler} T.,  {Dirsch} B.,  {Larsen} S.,  {Hilker} M.,   {Infante} L.,  2005,
  \mn@doi [\aap] {10.1051/0004-6361:20052705}, 439, 533

\bibitem[\protect\citeauthoryear{Rodriguez-Gomez et~al.,}{Rodriguez-Gomez
  et~al.}{2017}]{Rodriguez_Gomez_2017}
Rodriguez-Gomez V.,  et~al., 2017, \mn@doi [\mnras] {10.1093/mnras/stx305},
  467, 3083

\bibitem[\protect\citeauthoryear{{S{\'a}nchez-Janssen}
  et~al.,}{{S{\'a}nchez-Janssen} et~al.}{2019}]{Sanchez2019}
{S{\'a}nchez-Janssen} R.,  et~al., 2019, \mn@doi [\apj]
  {10.3847/1538-4357/aaf4fd}, \href
  {https://ui.adsabs.harvard.edu/abs/2019ApJ...878...18S} {878, 18}

\bibitem[\protect\citeauthoryear{Schaye et~al.}{Schaye
  et~al.}{2015}]{Schaye2015}
Schaye J.,  et~al., 2015, \mn@doi [\mnras] {10.1093/mnras/stu2058}, \href
  {http://adsabs.harvard.edu/abs/2015MNRAS.446..521S} {446, 521}

\bibitem[\protect\citeauthoryear{Seth et~al.}{Seth et~al.}{2014}]{Seth2014}
Seth A.~C.,  et~al., 2014, \mn@doi [\nat] {10.1038/nature13762}, \href
  {http://adsabs.harvard.edu/abs/2014Natur.513..398S} {513, 398}

\bibitem[\protect\citeauthoryear{Simmerer, Ivans, Filler, Francois, Charbonnel,
  Monier  \& James}{Simmerer et~al.}{2013}]{Simmerer2013}
Simmerer J.,  Ivans i.,  Filler D.,  Francois P.,  Charbonnel C.,  Monier R.,
  James G.,  2013, \mn@doi [\apjl] {10.1088/2041-8205/764/1/L7}, 764, L7

\bibitem[\protect\citeauthoryear{Spengler et~al.,}{Spengler
  et~al.}{2017}]{Spengler2017}
Spengler C.,  et~al., 2017, \mn@doi [\apj] {10.3847/1538-4357/aa8a78}, 849

\bibitem[\protect\citeauthoryear{{Springel}, {White}, {Tormen}  \&
  {Kauffmann}}{{Springel} et~al.}{2001}]{Springel2001}
{Springel} V.,  {White} S. D.~M.,  {Tormen} G.,   {Kauffmann} G.,  2001,
  \mn@doi [\mnras] {10.1046/j.1365-8711.2001.04912.x}, \href
  {https://ui.adsabs.harvard.edu/abs/2001MNRAS.328..726S} {328, 726}

\bibitem[\protect\citeauthoryear{{Thomas}, {Drinkwater}  \&
  {Evstigneeva}}{{Thomas} et~al.}{2008}]{Thomas2008}
{Thomas} P.~A.,  {Drinkwater} M.~J.,   {Evstigneeva} E.,  2008, \mn@doi
  [\mnras] {10.1111/j.1365-2966.2008.13543.x}, \href
  {https://ui.adsabs.harvard.edu/abs/2008MNRAS.389..102T} {389, 102}

\bibitem[\protect\citeauthoryear{{Trayford} et~al.,}{{Trayford}
  et~al.}{2015}]{Trayford2015}
{Trayford} J.~W.,  et~al., 2015, \mn@doi [\mnras] {10.1093/mnras/stv1461},
  \href {https://ui.adsabs.harvard.edu/abs/2015MNRAS.452.2879T} {452, 2879}

\bibitem[\protect\citeauthoryear{{Voggel}, {Hilker}  \& {Richtler}}{{Voggel}
  et~al.}{2016}]{Voggel2016}
{Voggel} K.,  {Hilker} M.,   {Richtler} T.,  2016, \mn@doi [\aap]
  {10.1051/0004-6361/201527070}, \href
  {https://ui.adsabs.harvard.edu/abs/2016A&A...586A.102V} {586, A102}

\bibitem[\protect\citeauthoryear{Voggel, Seth, Baumgardt, Mieske, Pfeffer  \&
  Rasskazov}{Voggel et~al.}{2019}]{Voggel_2019}
Voggel K.~T.,  Seth A.~C.,  Baumgardt H.,  Mieske S.,  Pfeffer J.,   Rasskazov
  A.,  2019, \mn@doi [\apj] {10.3847/1538-4357/aaf735}, 871, 159

\bibitem[\protect\citeauthoryear{{Wetzel}, {Cohn}  \& {White}}{{Wetzel}
  et~al.}{2009}]{Wetzel2009}
{Wetzel} A.~R.,  {Cohn} J.~D.,   {White} M.,  2009, \mn@doi [\mnras]
  {10.1111/j.1365-2966.2009.14424.x}, \href
  {https://ui.adsabs.harvard.edu/abs/2009MNRAS.395.1376W} {395, 1376}

\bibitem[\protect\citeauthoryear{{Zhang} et~al.,}{{Zhang}
  et~al.}{2015}]{Zhang2015}
{Zhang} H.-X.,  et~al., 2015, \mn@doi [\apj] {10.1088/0004-637X/802/1/30},
  \href {https://ui.adsabs.harvard.edu/abs/2015ApJ...802...30Z} {802, 30}

\bibitem[\protect\citeauthoryear{{Zhang} et~al.,}{{Zhang}
  et~al.}{2018}]{Zhang2018}
{Zhang} H.-X.,  et~al., 2018, \mn@doi [\apj] {10.3847/1538-4357/aab88a}, \href
  {https://ui.adsabs.harvard.edu/abs/2018ApJ...858...37Z} {858, 37}

\makeatother
\end{thebibliography}








\bsp	
\label{lastpage}
\end{document}